\documentclass{aastex}

\usepackage{graphicx}

\usepackage{emulateapj5}
\usepackage{times}

\makeatletter

\newenvironment{inlinefigure}{%
\def\@captype{figure}%
\noindent\begin{minipage}{0.999\linewidth}\begin{center}}
{\end{center}\end{minipage}\smallskip}
\makeatother

\usepackage{mathptmx}

\def\lax    {{_<\atop^{\sim}}}

\def\Mo     {{\rm M}_{\odot}}
\def\Lo     {{\rm L}_{\odot}}

\begin{document}

\slugcomment{accepted for publication by the {\em The Astrophysical Journal}}

\title{The Hot Gas Content of Low-Luminosity Early-Type Galaxies and the Implications
Regarding Supernova Heating and AGN Feedback}

\author{Laurence P. David, Christine Jones, William Forman, Iris Monica Vargas \& Paul Nulsen}
\affil{Harvard-Smithsonian Center for Astrophysics, 60 Garden St.,
Cambridge, MA 02138;\\ david@head.cfa.harvard.edu}

\shorttitle{\emph Galactic Winds in Ellipticals}

\begin{abstract}
We have analyzed Chandra observations of 18 low-luminosity early-type galaxies
with $L_B \lax 3 \times 10^{10} L_{\odot B}$.  Thermal emission from hot
gas with temperatures between 0.2 and 0.8~keV comprises
5-70\% of the total 0.5-2.0~keV emission
from these galaxies.   We find that the total X-ray luminosity from LMXBs
(resolved plus the power-law component of the unresolved emission) scales roughly
linearly with the K-band luminosity of the galaxies with a normalization
comparable to that found in more luminous early-type galaxies.
All of the galaxies in our sample are gas poor with gas masses much
less than that expected from the accumulation of stellar mass loss over the
life time of the galaxies.
The average ratio of gas mass to stellar mass in our sample is $M_{gas}/M_*=0.001$,
compared to more luminous early-type galaxies which typically have
$M_{gas}/M_*=0.01$.
The time required to accumulate the observed gas mass from stellar mass loss
in these galaxies is typically $3 \times 10^8$~yr.  Since the cooling
time of the gas is longer than the replenishment time, the gas cannot be condensing
out of the hot phase and forming stars, implying that the gas is most
likely being expelled from these galaxies in a wind. The one exception to this
is NGC4552, which is the most optically luminous galaxy in our sample and
has the highest gas content.  Using recent estimates of the Type Ia supernova rate
and AGN heating rate in early-type galaxies, we
find that, on average, heating by Type Ia supernovae
should exceed AGN heating in galaxies with $L_B \lax 3 \times 10^{10} L_{\odot B}$.
We also find that heating by Type Ia supernovae is energetically sufficient to
drive winds in these galaxies, even if the present Type Ia supernova rate is
overestimated by a factor of two or the present stellar mass loss rate
is underestimated by a factor of two.  There is significant scatter in
the gas properties of galaxies with comparable optical luminosities. Nearly
continuous heating by Type Ia supernovae alone cannot account for the
large scatter in observed gas mass.  We suggest that the scatter in
gas properties could arise from periodic AGN outbursts, during which time
AGN heating dominates over Type Ia supernova heating, or environmental factors,
such as a high pressure environment which would suppress the formation of
strong galactic winds.
\end{abstract}





\section{Introduction}

Heating by supernovae and active galactic nuclei (AGN) are thought to play 
important roles in galaxy formation, generating the observed correlation between
the bulge mass of a galaxy and the mass of the central supermassive black
hole, the upper mass cutoff of galaxies and reheating the gas in galaxies, 
groups and clusters.
In order to reconcile the predictions of the cold dark matter hierarchical clustering 
scenario with the observed luminosity function 
of galaxies and their mass-to-light ratios, it is thought that supernova feedback 
regulates star formation in galaxies less luminous than
about $3 \times 10^{10} L_{\odot B}$, while AGN feedback regulates 
star formation in more massive galaxies (e.g., White \& Frank 1991;
Bower et al. 2005; Croton et al. 2006).  Only Type II supernovae provide a 
natural feedback mechanism due to the short time delay between 
the onset of star formation and the first supernova. While Type Ia supernovae
(SN Ia) cannot provide a feedback mechanism that regulates concurrent 
star formation, SN Ia can play a significant role in the fate of gas shed 
by evolving stars in low-mass galaxies long after star formation ceases.


Einstein observations showed that the bulk of the X-ray
emission from optically luminous early-type galaxies arises
from hot gas in hydrostatic equilibrium (Forman, Jones \& Tucker 1985).
A harder X-ray component, more prevalent among low-luminosity 
early-type galaxies, was detected by ASCA and
assumed to arise from low-mass X-ray binaries (LMXBs) due to the
old stellar population in these systems (Kim, Fabbiano \& Trinchieri 1992).
Based on the analysis of ASCA data, White, Sarazin \& Kulkarni (2002)
found that the luminosity of the hard X-ray component
was more strongly correlated with the luminosity in globular
clusters rather than the total optical luminosity of the galaxies.
Chandra, with its superior angular resolution, has 
resolved populations of point sources in many early-type 
galaxies (e.g., Sarazin, Irwin \& Bregman 2000; Angelini, Lowenstein \& Mushotzky 2001;
Kraft et al. 2001; Blanton, Sarazin \& Irwin 2001) and
found that 20 to 80\% of the point sources reside in globular clusters 
(Sarazin et al. 2003).

While the hot gas content of X-ray luminous early-type galaxies
has been well studied by Chandra and previous X-ray telescopes,
little is known about the properties of the gas in less luminous 
early-type galaxies.  In low-luminosity early-type galaxies, the X-ray 
emission from LMXBs can exceed the thermal emission from 
hot gas, so it is imperative that the LMXBs be detected and excised from 
the analysis of the diffuse emission.
Theoretical models concerning the evolution 
of early-type galaxies show that for typical SN Ia rates, early-type 
galaxies fainter than
$M_B \approx -20$ should possess SN Ia driven galactic winds
(e.g., David, Forman \& Jones 1990, 1991). 
In addition to heating by SN Ia, AGN heating
may also expel a significant portion of the gas shed by evolving stars
in the shallow potential well of low-luminosity galaxies.  Chandra 
has detected X-ray cavities 
filled with radio emitting plasma and AGN driven shocks in many
elliptical galaxies and clusters 
(e.g., McNamara 

\begin{table*}[t]
\begin{center}
\caption{Low-Luminosity Early-Type Galaxy Sample}
\begin{tabular}{lccccccc}
\hline
Name & $D_L$ & $M_B$ & Type & $L_K$ & $\sigma_*$ & ObsIds & Exposure\\
       & (Mpc) & & & ($L_{\odot ,K}$) & (km~s$^{-1}$) & & (ksec) \\
\hline\hline
ESO 0428-G014 & 24.3 & -19.96 & -1.6 &  $3.62 \times 10^{10}$ & - & 4866 & 28.1 \\
NGC 821  & 24.1 & -20.19 & -5.0 &  $8.79 \times 10^{10}$ &  199.0 & 4006 4408 & 22.9 \\
NGC 1023 & 11.4 & -20.21 & -3.0 &  $8.44 \times 10^{10}$ &  204.0 & 4696 & 10.5 \\
NGC 1386 & 16.5 & -18.97 & -0.6 &  $3.47 \times 10^{10}$ &  166.0 & 4076 & 19.6 \\
NGC 1389 & 21.7 & -19.24 & -3.3 &  $3.91 \times 10^{10}$ &  132.0 & 4169 & 38.4 \\
NGC 2434 & 21.6 & -20.10 & -5.0 &  $7.38 \times 10^{10}$ &  205.0 & 2923 & 19.2 \\
NGC 2787 & 7.48 & -17.76 & -1.0 &  $1.46 \times 10^{10}$ &  194.0 & 4689 & 31.4 \\
NGC 3115 & 9.68 & -20.19 & -3.0 &  $8.95 \times 10^{10}$ &  266.0 & 2040 & 34.6 \\
NGC 3245 & 20.9 & -19.95 & -2.0 &  $5.92 \times 10^{10}$ &  220.0 & 2926 & 9.6  \\
NGC 3377 & 11.2 & -19.18 & -5.0 &  $2.75 \times 10^{10}$ &  131.0 & 2934 & 38.7 \\
NGC 3379 & 10.6 & -19.94 & -5.0 &  $7.31 \times 10^{10}$ &  201.0 & 1587 & 31.2  \\
NGC 3608 & 22.9 & -20.11 & -5.0 &  $5.24 \times 10^{10}$ &  204.0 & 2073 & 38.1 \\
NGC 4251 & 19.6 & -20.03 & -2.0 &  $6.67 \times 10^{10}$ &  119.0 & 4695 & 10.5 \\
NGC 4435 & 11.4 & -18.68 & -2.0 &  $2.38 \times 10^{10}$ &  157.0 & 2883 & 24.6 \\
NGC 4459 & 16.1 & -19.83 & -1.0 &  $7.58 \times 10^{10}$ &  172.0 & 2927 & 9.5 \\
NGC 4552 & 15.3 & -20.36 & -5.0 &  $1.05 \times 10^{11}$ &  261.0 & 2072 & 53.3 \\
NGC 4697 & 11.7 & -20.28 & -5.0 &  $9.12 \times 10^{10}$ &  165.0 & 4727 4728 & 72.6 \\
NGC 5866 & 15.3 & -20.10 & -1.0 &  $9.04 \times 10^{10}$ &  159.0 & 2879 & 28.0 \\
\hline
\end{tabular}
\end{center}

\noindent
Notes:  Galaxy name, luminosity distance ($D_L$), absolute blue magnitude ($M_B$), 
Galaxy Type in RC3,
K-band luminosity ($L_K$) based on $M_{K\odot}=3.33$ within the the same aperture 
as that used to extract the X-ray spectrum,
stellar velocity dispersion ($\sigma_*$), 
Chandra ObsIds used in the analysis, and the cleaned ACIS exposure time.
\end{table*}

\smallskip

\noindent
et al. 2000, Finoguenov \& Jones 2002, Fabian et al. 2003,
Blanton et al. 2003, Nulsen et al. 2005a, Nulsen et al. 2005b; 
Forman et al. 2005).  Based 
on the analysis of cavities found in a sample of galaxies, groups and clusters, 
Birzan et al. (2004) determined that the mechanical power of AGNs can be 
up to $10^4$ times their radio power. 
Thus, the thermal and dynamic properties of the hot gas 
in low-luminosity early-type galaxies are sensitive probes
of the present SN Ia rate and AGN activity in these
galaxies.

This paper is organized in the following manner.  In $\S$~2, we present
our low-luminosity early-type galaxy sample.  Section 3 contains 
the details of our Chandra data reduction.  The spectroscopic results for the
LMXBs and diffuse emission are discussed in 
$\S$~4 and the observed scaling between the total X-ray luminosity of the 
LMXB population and the K-band luminosity of the galaxies is 
presented in $\S$~5.  Sections 6 and 7 discuss the properties of the
hot gas, the inferred characteristics of the galactic winds, and
the implications regarding the Fe abundance in galaxies with 
galactic winds.  In $\S$~8, we examine the relative importance of
AGN and SN Ia heating in these galaxies.  Possible origins of the large scatter
in the observed gas properties among the galaxies are discussed in $\S$~9 and the 
main results of our paper are summarized in $\S$~10.

\section{Galaxy Sample}

Jones et al. (2006) have compiled a sample of early-type galaxies
that have been observed by the Chandra X-ray observatory. We 
initially extracted a sample of galaxies with 
$M_B$ between -18 and -20.4 (corresponding to $\rm{L}_B$ between
$3 \times 10^9$ and $3 \times 10^{10} ~ \rm{L}_{\odot B}$).
Galaxies with near-by companions and total ACIS exposure times less than 
5~ksec were then excluded from our sample.  After a preliminary analysis of the
Chandra data, we also excluded galaxies with fewer than 100 total net counts 
from further analysis.  Our final sample of 18 low-luminosity early-type 
galaxies is shown in Table 1 along with their optical 
and infrared properties and cleaned ACIS exposure times.  The 
absolute blue magnitudes given in Table 1 
are computed from $m_B(T^0)$ (RC3; de Vaucouleurs et al. 1991) 
and the distance moduli given in Tonry et al. (2001)
if available.  Otherwise, we use the corrected redshift in Faber et al. (1989), or
finally, the uncorrected 
redshifts.  We use $H_0$=70~km~s$^{-1}$~Mpc$^{-1}$ throughout.  
The K-band apparent magnitudes and luminosities given in Table 1 are 
derived directly from the 2MASS K-band images and are computed within the 
same aperture used to extract the X-ray spectrum from the galaxy.
The stellar velocity dispersion is obtained from Faber et al. (1989), if 
available, otherwise we use the average stellar
velocity dispersion from the literature as given in the LEDA catalog.

\section{Data Reduction}

All {\it Chandra} archival observations of the galaxies in our sample were reprocessed 
with CIAO 3.2.2 and CALDB 3.1.0 and screened
for background flares.  Fifteen of the galaxies in our sample were imaged on the 
S3 chip, two on the I3 chip and one on the S2 chip. Multiple exposures
of the same galaxy were combined into a single exposure.
For each galaxy, we generated a 0.3-6.0 keV image at full spatial resolution
for the chip on which the target was imaged.  
We then ran the CIAO {\it wavdetect} tool on each  
0.3-6.0~keV image with a detection threshold of $10^{-6}$ to 
generate the point source regions. Unlike more luminous early-type galaxies,
the galaxies in our sample have little hot gas, so the point source detection
efficiency is fairly uniform across the galaxies.
There is some area on the ACIS chip

\begin{table*}[t]
\begin{center}
\caption{Spectral Analysis of Combined Emission from Non-Nuclear Point Sources}
\begin{tabular}{lcccccccc}
\hline
Name & N & $N_{bg}$ & $L_{X,min}$ & $N_H(gal)$ & $\Gamma$ & $\chi^2$~/DOF & $F_{X,res}$ & $L_{X,res}$ \\
       & & & (erg~s$^{-1}$) & ($10^{20}$~cm$^{-2}$) & & & (erg~s$^{-1}$~cm$^{-2}$) & (erg~s$^{-1}$)  \\
\hline\hline
ESO 0428-G014 & 3 & 0.26 & $7.2 \times 10^{37}$ & 22.2 & 1.6              & -      & $4.19 \times 10^{-15}$ & $2.7 \times 10^{38}$ \\
NGC 821       & 6 & 0.38 & $1.2 \times 10^{38}$ & 6.24 & 1.6              & -      & $1.20 \times 10^{-14}$ & $7.8 \times 10^{38}$ \\
NGC 1023      &13 & 2.2  & $5.6 \times 10^{37}$ & 7.06 & 1.40 (1.06-1.74) & 4.2/5  & $4.32 \times 10^{-14}$ & $5.6 \times 10^{38}$ \\
NGC 1386      & 7 & 0.59 & $5.0 \times 10^{37}$ & 1.41 & 1.6              & -      & $2.18 \times 10^{-14}$ & $6.5 \times 10^{38}$ \\
NGC 1389      & 6 & 0.22 & $1.4 \times 10^{38}$ & 1.47 & 1.65 (1.33-1.99) & 9.7/8  & $3.02 \times 10^{-14}$ & $1.6 \times 10^{39}$ \\
NGC 2434      &11 & 0.29 & $7.0 \times 10^{37}$ & 12.2 & 1.44 (1.12-1.77) & 4.0/5  & $2.83 \times 10^{-14}$ & $1.5 \times 10^{39}$ \\
NGC 2787      &17 & 1.8  & $6.7 \times 10^{36}$ & 4.39 & 1.26 (1.09-1.43) & 16.5/15& $3.74 \times 10^{-14}$ & $2.2 \times 10^{38}$ \\
NGC 3115      &52 & 2.9  & $9.2 \times 10^{36}$ & 4.61 & 1.46 (1.40-1.52) & 70.0/66& $1.45 \times 10^{-13}$ & $1.5 \times 10^{39}$ \\
NGC 3245      & 3 & 0.36 & $1.4 \times 10^{38}$ & 2.11 & 1.6              & -      & $1.39 \times 10^{-14}$ & $6.4 \times 10^{38}$ \\
NGC 3377      &25 & 3.1  & $1.1 \times 10^{37}$ & 2.77 & 1.35 (1.26-1.45) & 49.4/37& $6.35 \times 10^{-14}$ & $8.4 \times 10^{38}$ \\
NGC 3379      &42 & 4.6  & $1.0 \times 10^{37}$ & 2.79 & 1.66 (1.61-1.72) & 93.1/75& $1.81 \times 10^{-13}$ & $2.2 \times 10^{39}$ \\
NGC 3608      &18 & 1.0  & $7.5 \times 10^{37}$ & 1.48 & 1.38 (1.08-1.70) & 2.3/6  & $2.10 \times 10^{-14}$ & $1.2 \times 10^{39}$ \\
NGC 4251      & 7 & 0.14 & $2.0 \times 10^{38}$ & 1.85 & 1.6              & -      & $2.35 \times 10^{-14}$ & $1.1 \times 10^{39}$ \\
NGC 4435      & 6 & 0.03 & $3.5 \times 10^{37}$ & 2.61 & 1.75 (1.41-2.11) & 6.6/5  & $2.25 \times 10^{-14}$ & $3.5 \times 10^{38}$ \\
NGC 4459      & 5 & 0.55 & $9.0 \times 10^{37}$ & 2.68 & 1.6              & -      & $1.34 \times 10^{-14}$ & $3.7 \times 10^{38}$ \\
NGC 4552      &89 & 5.0  & $1.8 \times 10^{37}$ & 2.56 & 1.36 (1.32-1.40) &160/128 & $1.87 \times 10^{-13}$ & $4.9 \times 10^{39}$ \\
NGC 4697      &79 & 8.2  & $7.4 \times 10^{36}$ & 2.14 & 1.52 (1.48-1.57) &130/122 & $1.42 \times 10^{-13}$ & $2.1 \times 10^{39}$ \\
NGC 5866      &29 & 1.3  & $2.9 \times 10^{37}$ & 1.47 & 1.75 (1.62-1.89) & 19.2/21& $5.79 \times 10^{-14}$ & $1.5 \times 10^{39}$ \\
\hline
\end{tabular}
\end{center}

\noindent
Notes:  Galaxy name, number of detected non-nuclear point sources (N), expected number 
of background sources ($N_{bg}$), luminosity of a source in the galaxy with a flux equal 
to the $3 \sigma$ point source detection limit($L_{X,min}$), galactic 
hydrogen column density
($N_H(gal)$), best-fit power-law index ($\Gamma$) and 90\% confidence limit,
$\chi^2$ per degrees of freedom, and the background subtracted
(both the diffuse and AGN components of the X-ray background) 
0.5-2.0~keV flux ($F_{X,res}$) and 
luminosity ($L_{X,res}$) of the combined emission from the detected non-nuclear 
point sources. 
The flux and luminosity for spectra with fewer than 100 counts were calculated
assuming $\Gamma=1.6$ and galactic absorption.
\end{table*}

\smallskip

\noindent
beyond the 
$D_{25}$ optical isophote for all the galaxies in our sample so we can obtain a 
local background.  This is essential for the detection of 0.3-0.8~keV gas,
since the X-ray background below 0.8~keV is subject to galactic emission and 
charge exchange emission from solar wind particles in the Earth's magnetosphere
(Wargelin et al. 2004).  There is also a spatial gradient in the depth of the 
contaminant on the ACIS filters which affects the spectral shape of the 
background below 1~keV.  We therefore exclude data in the 
bottom and top 200 rows on ACIS-S and beyond a $5^{\prime}$ radius 
from the center of ACIS-I from further analysis.  Observations of the ACIS external 
calibration source show that the depth of the contaminant is fairly 
uniform in the remaining 
regions\footnote{http://cxc.harvard.edu/ciao/why/acisqedeg.html}.  
Our screening procedure ensures an accurate subtraction of the 
soft X-ray background which is vital for determining the gas properties
in these galaxies. 

\section{Spectral Analysis}

Using the point source regions generated by {\it wavdetect} along with
the $D_{25}$ optical isophote,  we extracted three spectra for each galaxy:
1) a spectrum containing the combined emission from all detected non-nuclear point sources
within the $D_{25}$ isophote,
2) a spectrum of the unresolved emission within $D_{25}$ (i.e., the emission 
outside of the detected source regions), and 3) a background spectrum 
from the unresolved emission beyond the $D_{25}$ isophote.
A point source is identified as an AGN if it is located within $2.0^{\prime\prime}$
of the centroid of the galaxy as determined from the 2MASS K-band image. 
None of our galaxies has more than one point source within this region.
The spectra for the detected sources and the unresolved emission 
were binned to a minimum
of 20 counts per bin and corresponding photon weighted response and area files were 
generated using the CIAO tasks {\it mkacisrmf} and {\it mkwarf}.  

\subsection{Point Source Population}

The background subtracted (i.e., the unresolved component of the
X-ray background) spectra of the combined emission from the detected 
non-nuclear point sources with more than 100 net counts were fitted to an absorbed power-law model 
with the absorption fixed at the galactic value. 
Our analysis shows that the emission from the binary populations is well fitted by
a power-law spectrum with $\Gamma \approx 1.6$ (see Table 2).
This result is consistent with the spectral analysis of the combined emission 
from LMXBs in a sample of 15 early-type galaxies analyzed by Irwin, Athey \& Bregman (2003).
In cases were the combined emission from the detected point sources results in fewer than 100 net 
counts, the fluxes are derived assuming galactic absorption
and $\Gamma = 1.6$.  To estimate the number of background point 
sources and integrated flux, we used the 0.5-2.0~keV luminosity function derived from the 
Chandra Deep Field-South (Rosati et al. 2002) and the $3 \sigma$ point source 
detection threshold in each observation.  The total number of detected non-nuclear point 
sources and the expected number of background AGN
above 

\begin{table*}[t]
\begin{center}
\caption{Results of Fitting the Diffuse Emission to an Absorbed Power-Law}
\begin{tabular}{lcc}
\hline
Name & $\Gamma$ & $\chi^2$~/DOF \\
\hline\hline
ESO 0428-G014 & 2.8 (2.0-3.6) & 11.9/10 \\
NGC 821       & 1.7 (1.0-2.4) & 17.6/19 \\
NGC 1023      & 2.5 (2.0-3.0) & 13.2/13 \\
NGC 1386      & 3.0 (2.8-3.2) & 111/45 \\
NGC 1389      & 1.8 (1.1-2.6) & 15.7/14 \\
NGC 2434      & 2.8 (2.6-3.0) & 125/38 \\
NGC 2787      & 2.6 (2.3-2.8) & 20.1/13 \\
NGC 3115      & 1.9 (1.8-2.0) & 43.4/29 \\
NGC 3245      & 3.0 (2.7-3.3) & 19.6/12 \\
NGC 3377      & 2.5 (2.2-2.8) & 15.9/15 \\
NGC 3379      & 2.0 (1.8-2.2) & 39.4/32 \\
NGC 3608      & 3.4 (3.2-3.6) & 41.0/14 \\
NGC 4251      & 2.0 (1.5-2.4) & 5.6/8 \\
NGC 4435      & 3.0 (2.7-3.2) & 39.5/38 \\
NGC 4459      & 1.9 (1.6-2.3) & 50.4/13 \\
NGC 4552      & 2.3 (2.1-2.4) & 4062/133 \\
NGC 4697      & 2.8 (2.7-2.9) & 344/84 \\
NGC 5866      & 2.6 (2.5-2.7) & 183/67 \\
\hline
\end{tabular}
\end{center}

\noindent
Notes:  Galaxy name, best-fit power-law index ($\Gamma$) and 90\% confidence limit, 
and the $\chi^2$ per degrees of freedom.
\end{table*}

\smallskip

\noindent
the $3 \sigma$ detection limit are shown in Table 2.  The resulting 
background subtracted (both unresolved and AGN) 0.5-2.0~keV flux and luminosity for the combined 
point source emission in each galaxy are given in Table 2.
Due to the small angular extent of the galaxies, variations in the number
of background AGN do not have a significant effect on the derived fluxes. 
There are also very few photons from the diffuse background component 
within the source regions.  
The primary uncertainty in the derived X-ray flux is the
uncertainty in the spectral model.  For $\Gamma$ between 1.2 and 2.0,
the estimated 0.5-2.0~keV flux varies by 25\%.


\subsection{Diffuse Emission}

The unresolved diffuse emission in each observation must contain some emission
from LMXBs with fluxes below the point source detection limit.
The low gas temperature ($kT \approx 0.3-0.6$~keV) in these 
galaxies makes it easier to separate the thermal and power-law components
compared to more X-ray luminous early-type galaxies which typically have gas 
temperatures around 1~keV.  
We first fit the diffuse emission with an absorbed 
power-law model.  Except for N1389, N821 and N4251, the best fit index is significantly
steeper than that obtained from fitting the combined emission from the detected
point sources (see Table 3), indicating the presence of a diffuse soft component.
We then fit the diffuse spectra with an absorbed power-law 
plus MEKAL model with the absorption frozen at the galactic value and 
the abundance of heavy elements frozen at the solar value (see the results in Table 4).
In all but three cases (N1389, N821 and N4251), adding a thermal component
improved the fit at greater than the 95\% confidence level based on a F-test.
The resulting best-fit power-law indices in the two component model
are consistent with those derived 
from fitting the combined emission from the detected point sources.
This shows that the spectral index of LMXBs does not vary significant 
with luminosity, which is consistent with Chandra observations of the 
lowest luminosity LMXBs in our Galaxy (Wilson et al. 2003).
For N1389, N821 and N4251, we determined the 90\% upper limit on the 
thermal flux by increasing the normalization of the MEKAL model
until the resulting $\chi^2$ increased by 2.71 compared to the 
best-fit with a pure power-law model.


Besides emission from hot gas, there are other potential sources for
the unresolved soft X-ray emission, including:
supersoft sources, M stars and RS CVn systems. 
Pellegrini \& Fabbiano (1994) showed that both K-M main-sequence stars,
due to the reduced rotation rate in old stars,
and Rs CVn systems are unlikely to produce a significant 
component of the soft X-ray emission in early-type galaxies.
Supersoft sources are characterized 
by black body emission with $kT < 100$~eV.  Supersoft sources are probably a combination
of several types of objects, including: postnovae hot white dwarf, symbiotics, pre-white
dwarfs in planetary nebulae, accreting white dwarfs
and possibly accreting intermediate mass black holes (Kahabka \& van den Heuvel 1997).
A scatter plot of gas temperature vs. stellar velocity dispersion
for our galaxy sample is shown in Fig. 1.  The solid line 
in Fig. 1 is the relation found by 
O'Sullivan, Ponman \& Collins (2003) 
for a sample of more X-ray luminous early-type galaxies with typical
velocity dispersions between 250 and 350~km~s$^{-1}$.
While there is significant scatter in Fig. 1, the relation between gas temperature
and velocity dispersion in our sample of galaxies is in reasonable agreement with 
that observed in hotter systems providing strong evidence that the 
soft X-ray emission in our sample arises from hot gas.
Gas shed by evolving stars thermalizes at the temperature associated
with the stellar velocity dispersion of the stars ($T_* = \mu m_p \sigma_*^2 / k$),
which is shown as a dashed line in Fig. 1. 
The ratio of energy per unit mass in the stars to that in the gas,
$\beta_{spec} = T_* / T_{gas}$, varies from 0.3 to 1 in our sample, which is 
comparable to values found in more X-ray luminous early-type galaxies, giving further
support to a gaseous origin for the soft unresolved emission.

\begin{table*}[t]
\begin{center}
\caption{Results of Fitting the Diffuse Emission to an Absorbed Power-Law Plus Mekal Model}
\begin{tabular}{lcccccccc}
\hline
Name & $\Gamma$ & kT & $\chi^2$~/DOF & $L_{X,unres}$ & $L_{X,LMXB}$ & $f_{unres}$ & $L_{X,gas}$ & $f_{gas}$ \\
       & & (keV) & & (erg~s$^{-1}$) & (erg~s$^{-1}$) & (\%) & (erg~s$^{-1}$) & (\%) \\
\hline\hline
ESO 0428-G014 & 0.8 (0.4-2.0) & 0.60 (0.34-0.83) & 5.9/8   & $2.5 \times 10^{38}$ & $5.2 \times 10^{38}$ & 48 & $9.5 \times 10^{38}$ & 65 \\
NGC 821       & 1.7           & 0.4              & 14.6/17 & $1.1 \times 10^{39}$ & $1.9 \times 10^{39}$ & 58 & $< 7.0 \times 10^{38}$ & $<$ 27 \\
NGC 1023      & 2.5 (1.2-3.5) & 0.78 (0.34-1.10) & 6.2/11  & $4.6 \times 10^{38}$ & $1.0 \times 10^{39}$ & 46 & $5.3 \times 10^{38}$ & 35 \\
NGC 1386      & 1.7 (0.6-2.7) & 0.29 (0.25-0.32) & 68.0/43 & $8.1 \times 10^{38}$ & $1.5 \times 10^{39}$ & 54 & $2.4 \times 10^{39}$ & 61 \\
NGC 1389      & 1.8           & 0.4              & 14.7/12 & $4.9 \times 10^{38}$ & $2.1 \times 10^{39}$ & 23 & $< 5.7 \times 10^{38}$ & $<$ 21 \\
NGC 2434      & 1.9 (1.4-2.5) & 0.51 (0.44-0.58) & 26.8/36 & $1.4 \times 10^{39}$ & $3.0 \times 10^{39}$ & 47 & $6.0 \times 10^{39}$ & 67 \\
NGC 2787      & 2.0 (0.8-3.1) & 0.35 (0.27-0.60) & 6.1/11  & $1.8 \times 10^{38}$ & $4.0 \times 10^{38}$ & 45 & $2.1 \times 10^{38}$ & 34 \\
NGC 3115      & 1.6 (1.4-1.7) & 0.52 (0.36-0.74) & 23.7/28 & $5.5 \times 10^{38}$ & $2.0 \times 10^{39}$ & 27 & $3.2 \times 10^{38}$ & 14 \\
NGC 3245      & 2.4 (1.4-3.8) & 0.32 (0.22-0.54) & 11.5/10 & $7.8 \times 10^{38}$ & $1.4 \times 10^{39}$ & 56 & $1.1 \times 10^{39}$ & 44 \\
NGC 3377      & 1.9 (0.9-2.7) & 0.21 (0.12-0.38) & 10.6/13 & $2.0 \times 10^{38}$ & $1.0 \times 10^{39}$ & 20 & $1.6 \times 10^{38}$ & 14 \\
NGC 3379      & 1.9 (1.5-2.4) & 0.65 (0.20-1.1)  & 34.8/30 & $6.6 \times 10^{38}$ & $2.8 \times 10^{39}$ & 24 & $1.2 \times 10^{38}$ & 4 \\
NGC 3608      & 2.4 (1.9-3.1) & 0.35 (0.29-0.47) & 19.2/12 & $1.4 \times 10^{39}$ & $2.6 \times 10^{39}$ & 54 & $2.7 \times 10^{39}$ & 51 \\
NGC 4251      & 2.0           & 0.4              &  4.9/6  & $9.2 \times 10^{38}$ & $2.0 \times 10^{39}$ & 46 & $< 7.8 \times 10^{38}$ & $<$ 28 \\
NGC 4435      & 2.4 (1.5-4.0) & 0.39 (0.28-0.85) & 28.7/36 & $5.2 \times 10^{38}$ & $8.7 \times 10^{38}$ & 60 & $4.8 \times 10^{38}$ & 36 \\
NGC 4459      & 0.9 (0.4-1.7) & 0.42 (0.32-0.52) & 18.4/11 & $4.0 \times 10^{38}$ & $7.7 \times 10^{38}$ & 52 & $1.8 \times 10^{39}$ & 70 \\
NGC 4552      & 1.9 (1.8-2.0) & 0.54 (0.51-0.56) & 179/131 & $3.4 \times 10^{39}$ & $8.3 \times 10^{39}$ & 41 & $1.8 \times 10^{40}$ & 68 \\
NGC 4697      & 1.6 (1.0-2.2) & 0.33 (0.31-0.36) & 92.2/81 & $1.8 \times 10^{38}$ & $2.3 \times 10^{39}$ & 8  & $2.1 \times 10^{39}$ & 47 \\ 
NGC 5866      & 1.1 (0.9-1.2) & 0.32 (0.30-0.35) & 84.3/65 & $6.2 \times 10^{38}$ & $2.1 \times 10^{39}$ & 29 & $2.2 \times 10^{39}$ & 51 \\
\hline
\end{tabular}
\end{center}

\noindent
Notes:  Galaxy name, best-fit power-law index ($\Gamma$) and temperature (kT) along with 
their 90\% confidence limits, $\chi^2$ per degrees of freedom, X-ray luminosity of the 
power-law component of the diffuse emission ($L_{X,unres}$), total X-ray luminosity 
of the LMXBs ($L_{X,LMXB}=L_{X,res}+L_{X,unres}$), fraction of the power-law 
component that is unresolved ($f_{unres}$), X-ray luminosity of the thermal component 
of the diffuse emission ($L_{X,gas}$), and the fraction of the total 
luminosity arising from gas ($f_{gas}$).  All luminosities are given in the 
0.5-2.0~keV bandpass. The values of $L_{X,gas}$ and $f_{gas}$ for NGC821, NGC1389,
and NGC4251 correspond to 90\% upper limits.
\end{table*}

\smallskip

\begin{inlinefigure}
\center{\includegraphics*[width=0.90\linewidth,bb=10 142 570 700,clip]{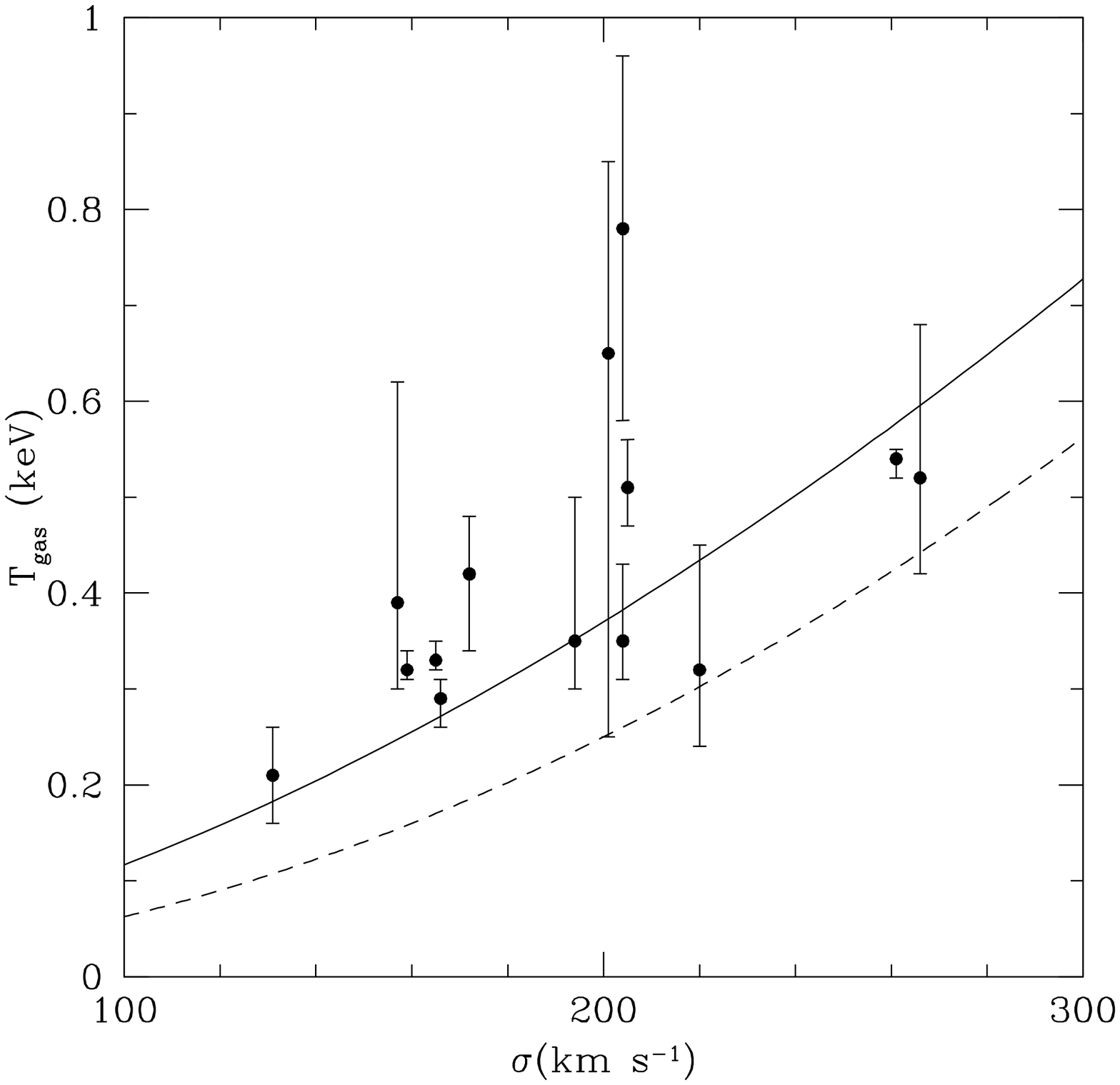}}
\caption{Scatter plot of the best-fit temperature of the diffuse emission
vs. the stellar velocity dispersion. The solid line is
the relation between $T_{gas}$ and $\sigma_*$ found by O'Sullivan et al. (2003)
for a sample of more X-ray luminous early-type galaxies
and the dashed line is the temperature associated with the stellar velocity dispersion
in the galaxies.}
\end{inlinefigure}

\smallskip

We also can use the Chandra data directly to place limits on the 
contribution to the diffuse emission from supersoft sources.
Chandra has detected supersoft sources in many galaxies 
(Sarazin, Irwin \& Bregman 2000; Pence et al. 2001; Swartz et al. 2002; 
Di Stefano \& Kong 2004) and there appears to be a general trend that 
supersoft sources are more common in the spiral arms of late-type galaxies
compared to bulge dominated systems.
Sarazin et al. (2000) identified 3 supersoft sources in NGC 4697 based on 
hardness ratios.  Di Stefano \& Kong (2004) fitted the combined emission
from the 3 supersoft sources in NGC 4697 to an absorbed black body
model and obtained a temperature of $kT=80$~eV
and a 0.3-7.0~keV flux of $1.89 \times 10^{-14}$~erg~cm$^{-2}$~s$^{-1}$.
Based on this spectral model, and converting between bandpasses, this 
gives a 0.5-2.0~keV flux of $6.0 \times 10^{-15}$~erg~cm$^{-2}$~s$^{-1}$,
which is only 2\% of the total resolved 0.5-2.0~keV flux of NGC4697 given in Table 2.  
Adding an 80~eV black body component in the spectral analysis of the diffuse emission does not 
significantly improve the fit in any of the galaxies in our sample.  For NGC 4697,
the 90\% uppper limit on the 0.5-2.0~keV flux of an 80~eV black body component
is $6.0 \times 10^{-15}$~erg~cm$^{-2}$~s$^{-1}$, which is twice the flux
of the resolved supersoft sources, but only 4\% of the flux in the thermal
component of the unresolved emission.

\section{Scaling Between ${l}_{x,lmxb}$ and the K-Band Luminosity}

\noindent
Earlier studies based on {\it Einstein} and {\it ROSAT} observations
showed that $L_x \propto L_B^{2}$ for early-type galaxies more luminous 
than $L_B \approx 3 \times 10^{10} \Lo$ and $L_x \propto L_B$ 
for less luminous galaxies 
(Forman, Jones \& Tucker 1985; Trinchieri \& Fabbiano 1985; 
O'Sullivan, Forbes \& Ponman  1991).  The different slopes were 
thought to arise from the competition between the two primary

\begin{inlinefigure}
\center{\includegraphics*[width=0.90\linewidth,bb=10 142 570 700,clip]{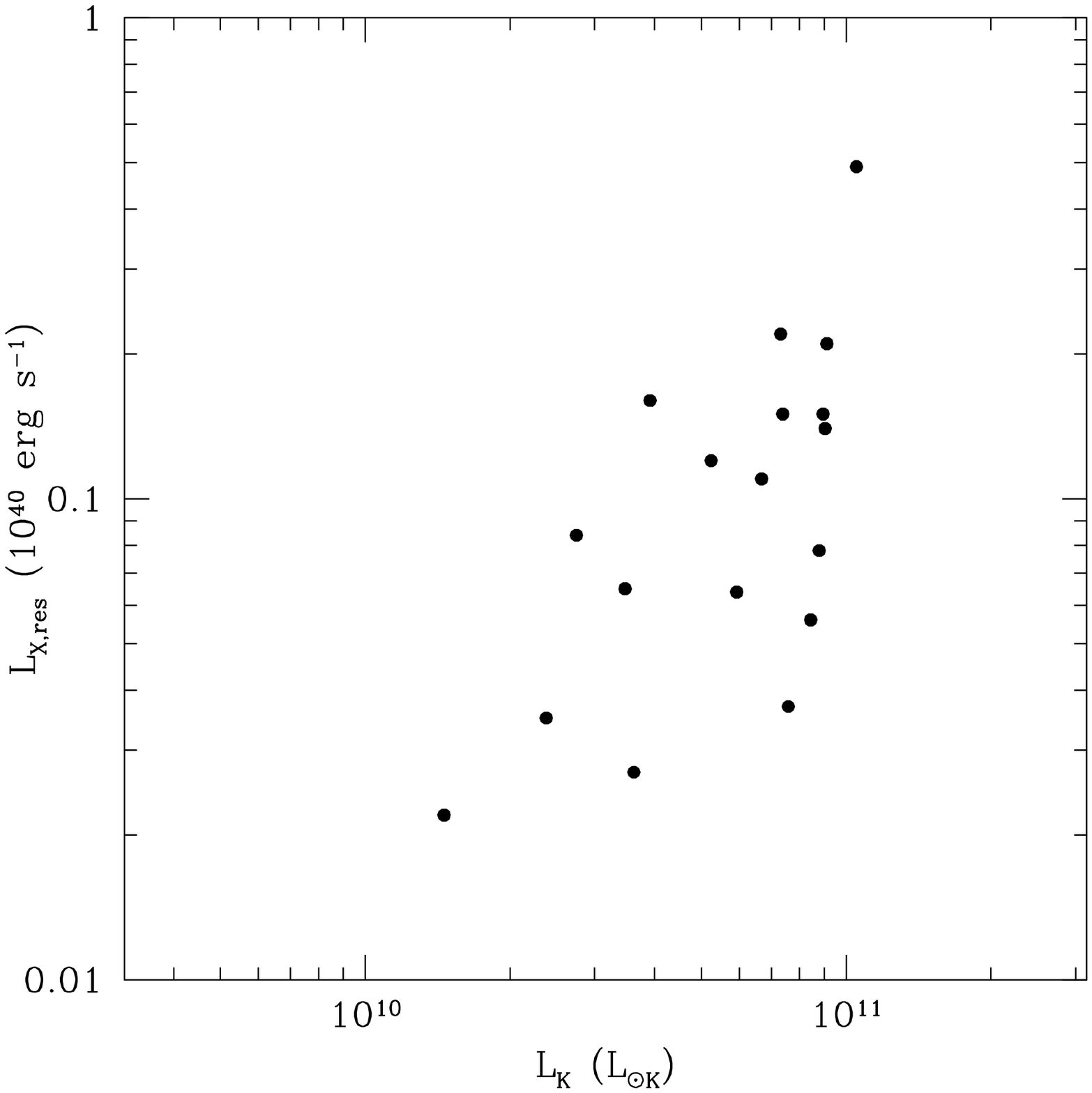}}  
\caption{Scatter plot of the combined 0.5-2.0~keV luminosity of the resolved LMXBs 
vs. the K-band luminosity of the galaxies.}
\end{inlinefigure}

\smallskip

\noindent
sources of X-rays in early-type galaxies, with emission from LMXBs 
dominating in low luminosity galaxies and thermal emission from hot 
gas dominating in more luminous galaxies.
Observations of early-type galaxies by Chandra are able to study these
two components separately.
With the availability of 2MASS data, it is more appropriate to use
$L_K$ instead of $L_B$ since it is a better proxy for stellar mass.
The combined 0.5-2.0~keV luminosity of only the resolved LMXBs
is plotted against the K-band luminosity of the galaxies in Fig. 2
and the total 0.5-2.0~keV luminosity of the LMXBs 
($L_{LMXB} = L_{X,res} + L_{X,unres}$)
is plotted against the K-band luminosity of the galaxies in Fig. 3.  Due to the
range in sensitivity among the observations (see Table 2), there is 
significant scatter in Fig. 2.  By combining the power-law component of the 
unresolved emission with the resolved LMXB emission, the scatter is significantly 
reduced in Fig. 3 and there is an overall linear trend between $L_{LMXB}$ and $L_K$.  
The most X-ray luminous galaxy in Figs. 2 and 3 is NGC4552, which is discussed
more extensively in $\S 9$.
Fitting a power-law to the data in Fig. 3 yields a best-fit of:

$$L_{LMXB} = 3.7 \times 10^{38} \left( {L_K} \over {10^{10} L_{\odot~K}} \right)^{0.81}~\rm{erg~s}^{-1} \eqno(1)$$

\noindent
which is shown as a solid line in Fig. 3.  
Several studies have shown that the spatial distribution of LMXBs follows
the light in early-type galaxies (Sarazin et al. 2001; 
Gilfanov 2002; Jordan et al. 2004),
so our estimate of $L_{LMXB}/L_K$ should be valid within any aperture.


Both Colbert et al. (2004) and Kim \& Fabbiano (2004) 
estimated the ratio of $L_{LMXB}$ to $L_K$ for samples 
of more X-ray luminous early-type galaxies. In these studies,
$L_{LMXB}$ was computed in the 0.3-8.0~keV bandpass assuming a count
rate conversion factor appropriate for power-law spectra with $\Gamma=1.7-1.8$ 
and galactic absorption.  The difference between the spectral index used in 
our work and these studies does not make a significance difference 
on the derived luminosities.  However, their 0.3-8.0~keV luminosities should be 
3 times our 0.5-2.0~keV luminosities.  Both studies only included
the emission from 

\begin{inlinefigure}
\center{\includegraphics*[width=0.90\linewidth,bb=10 142 570 700,clip]{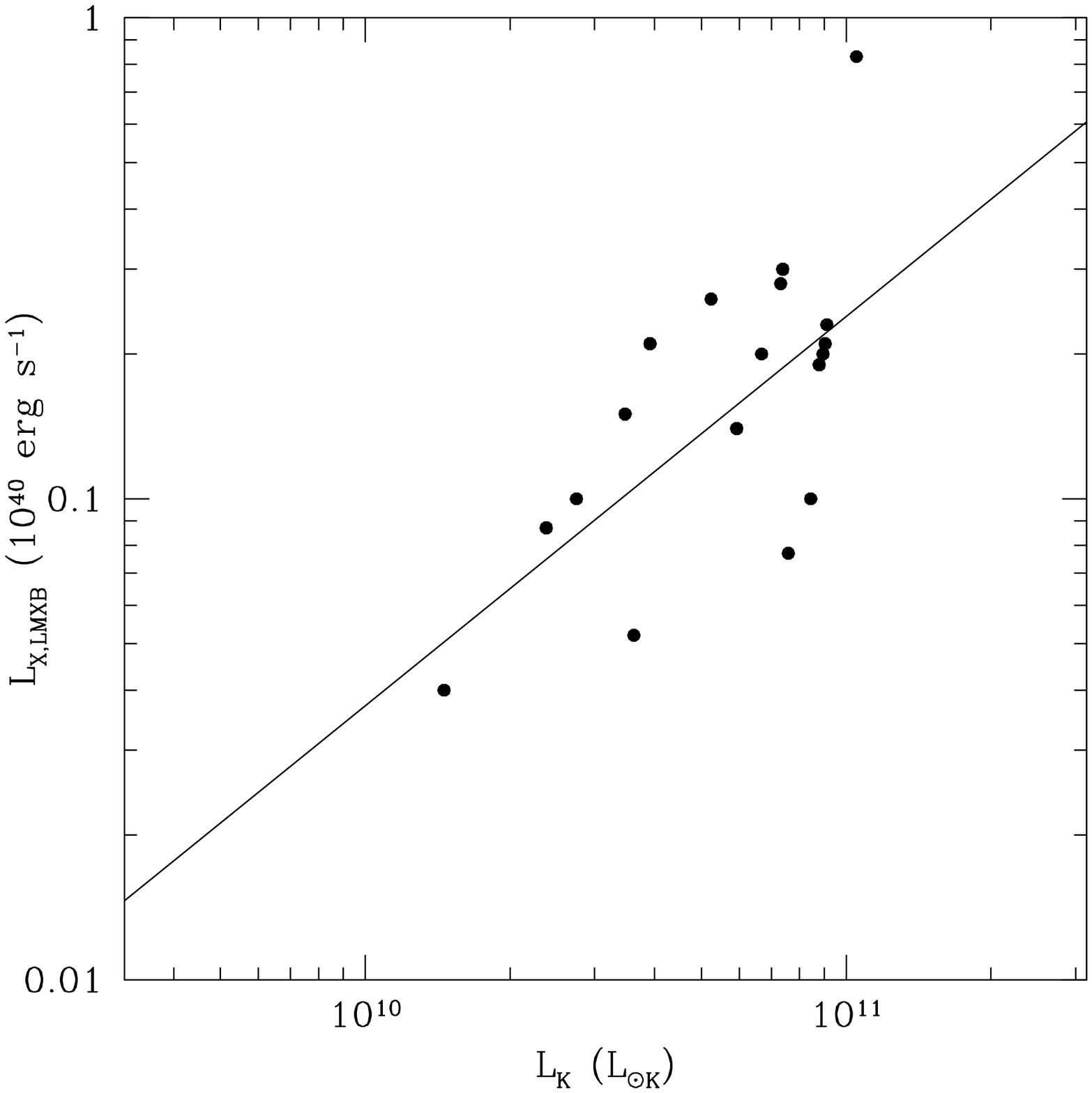}}  
\caption{Scatter plot of the total 0.5-2.0~keV luminosity of the LMXBs (resolved
plus the power-law component of the unresolved emission) vs. the K-band luminosity of the galaxies.}
\end{inlinefigure}

\smallskip

\noindent
non-nuclear point sources within the $D_{25}$ isophote.  
Kim \& Fabbiano computed $L_{LMXB}$ by integrating over the galaxy's LMXB luminosity function.
Colbert et al. derived a relation of $L_{LMXB}=1.3 \times 10^{-4}~L_K$,
where $L_K=\nu L_{\nu}$.  From the 2MASS 
documentation\footnote{http://www.ipac.caltech.edu/2mass},
$\nu F_{\nu} (m_K=0) = 9.27 \times 10^{-7}$~erg~cm$^{-2}$~s$^{-1}$. Rewriting their
expression in terms of $L_{\odot~K}$, and adjusting for bandpass 
differences, gives a coefficient in eq. (1) of $2.3 \times 10^{38}$~erg~s$^{-1}$, 
which is 50\% lower than our estimate.
After correcting for bandpass differences, the ratio of $L_{LMXB}$ to $L_K$
in Kim \& Fabbiano is a factor of 1.8 higher than our result.  Thus,
within a factor of two, we find the same ratio of $L_{LMXB}$ to $L_K$
in our sample of low-luminosity early-type galaxies as that found
in more X-ray luminous systems.

Previous studies have suggested that $L_{X,LMXB}$ is better correlated with 
the total luminosity in globular clusters (White, Sarazin \& Kulkarni 2002) 
and that the dispersion in the $L_{X,LMXB}$-$L_K$ relation is correlated
with the specific frequency of globular clusters, $S_N$ (Kim \& Fabbiano 2004).
A search of the literature only uncovered published values of $S_N$ for half
of the galaxies in our sample, and most of these estimates have large errors, so 
we cannot investigate the dependence of $L_{X,LMXB}$ on $S_N$ in our sample.

\section{The Hot Gas Content}

There is significant scatter in $L_{X,gas}$ compared to $L_{X,LMXB}$ 
with a factor of 100 variation in $L_{X,gas}$ for a given $L_K$ (see Fig. 4).
The ratio of $L_{X,gas}$ to $L_{X,LMXB}$ also shows significant scatter 
and varies from 0.04 to 2.2 (see Fig. 5).  
We also divided the sample into ellipticals ($T < -3$) and S0s ($T> -3.0$).
The average $L_{X,gas}$ is essentially the same for ellipticals and S0s, 
but the ellipticals have a greater dispersion (see Fig. 4).  Early-type
galaxies can be classified as "core" or "cuspy" depending on their inner stellar
surface brightness profile.  Based on ROSAT observations, Pellegrini (2005) found that 
for a given optical luminosity, core early-type galaxies are more X-ray luminous

\begin{table*}[t]
\begin{center}
\caption{Hot Gas Content}
\begin{tabular}{lcccccc}
\hline
Name & $R_m$ & $n_{e,0}$ & $M_{gas}$ & $\dot M_*$ & $t_g$ & $t_c$  \\
     & (kpc) &  (cm$^{-3}$) & ($M_{\odot}$) & ($M_{\odot}$~yr$^{-1}$) & (yr) & (yr) \\
\hline\hline
ESO 0428-G014 & 4.8 & $9.5 \times 10^{-3}$ & $2.0 \times 10^7$ & 0.065 & $3.1 \times 10^8$ & $2.2 \times 10^9$ \\
NGC 821       & 7.0 & $< 1.3 \times 10^{-2}$ & $< 5.0 \times 10^7$ & 0.158 & $< 3.2 \times 10^8$ & - \\
NGC 1023      & 6.9 & $5.6 \times 10^{-3}$ & $2.2 \times 10^7$ & 0.152 & $1.4 \times 10^8$ & $5.4 \times 10^9$ \\
NGC 1386      & 5.7 & $1.8 \times 10^{-2}$ & $4.9 \times 10^7$ & 0.062 & $7.9 \times 10^8$ & $7.0 \times 10^8$ \\
NGC 1389      & 5.7 & $< 1.4 \times 10^{-2}$ & $< 3.9 \times 10^7$ & 0.070 & $< 5.5 \times 10^8$ & - \\
NGC 2434      & 7.5 & $2.1 \times 10^{-2}$ & $9.3 \times 10^7$ & 0.133 & $7.0 \times 10^8$ & $1.4 \times 10^9$ \\
NGC 2787      & 3.9 & $5.6 \times 10^{-3}$ & $8.2 \times 10^6$ & 0.026 & $3.1 \times 10^8$ & $1.6 \times 10^9$ \\
NGC 3115      & 5.8 & $5.5 \times 10^{-3}$ & $1.6 \times 10^7$ & 0.161 & $9.9 \times 10^7$ & $4.6 \times 10^9$ \\
NGC 3245      & 7.3 & $1.1 \times 10^{-2}$ & $4.6 \times 10^7$ & 0.107 & $4.3 \times 10^8$ & $1.6 \times 10^9$ \\
NGC 3377      & 6.5 & $4.7 \times 10^{-3}$ & $1.7 \times 10^7$ & 0.049 & $3.5 \times 10^8$ & $2.6 \times 10^9$ \\
NGC 3379      & 7.7 & $3.1 \times 10^{-3}$ & $1.4 \times 10^7$ & 0.132 & $1.1 \times 10^8$ & $1.3 \times 10^{10}$ \\
NGC 3608      & 9.3 & $1.4 \times 10^{-2}$ & $9.0 \times 10^7$ & 0.094 & $9.6 \times 10^8$ & $1.4 \times 10^9$ \\
NGC 4251      & 6.8 & $< 8.4 \times 10^{-3}$ & $< 3.2 \times 10^7$ & 0.120 & $< 2.7 \times 10^8$ & - \\
NGC 4435      & 4.0 & $7.7 \times 10^{-3}$ & $1.2 \times 10^7$ & 0.043 & $2.8 \times 10^8$ & $1.4 \times 10^9$ \\
NGC 4459      & 7.0 & $1.2 \times 10^{-2}$ & $4.7 \times 10^7$ & 0.136 & $3.4 \times 10^8$ & $1.6 \times 10^9$ \\
NGC 4552      & 10.7 & $3.2 \times 10^{-2}$ & $2.5 \times 10^8$ & 0.189 & $1.3 \times 10^9$ & $1.3 \times 10^9$ \\
NGC 4697      & 9.1 & $1.3 \times 10^{-2}$ & $8.2 \times 10^7$ & 0.164 & $5.0 \times 10^8$ & $1.5 \times 10^9$ \\
NGC 5866      & 6.7 & $1.6 \times 10^{-2}$ & $5.9 \times 10^7$ & 0.163 & $3.6 \times 10^8$ & $1.0 \times 10^9$ \\
\hline
\end{tabular}
\end{center}

\noindent
Notes:  Galaxy name, the radius of a circular aperture which has the same 
area as the region used to extract the diffuse X-ray spectrum
($R_m$), central electron number density ($n_{e,0}$), gas mass ($M_{gas}$) within 
$R_m$, stellar mass loss rate ($\dot M_*)$ estimated from
$L_K$ in Table 1, the time to accumulate the observed gas mass given the
present stellar mass loss rate ($t_g$), and the average radiative cooling time of the 
gas ($t_c$).
\end{table*}

\smallskip

\begin{inlinefigure}
\center{\includegraphics*[width=0.90\linewidth,bb=10 142 570 700,clip]{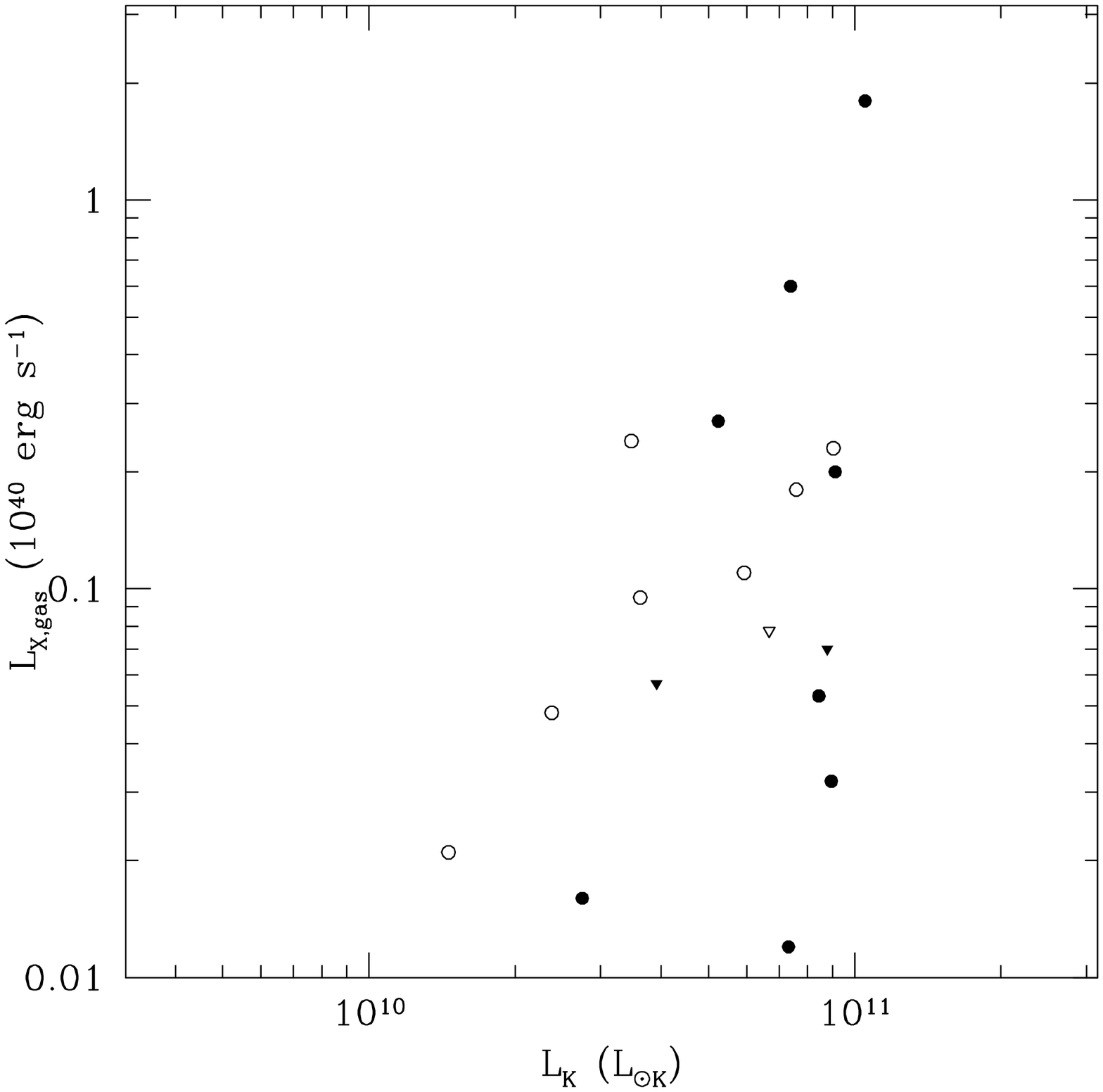}}  
\caption{Scatter plot of the 0.5-2.0~keV luminosity of the thermal component of the 
diffuse emission ($L_{X,gas}$) vs. the K-band luminosity of the galaxies. Triangles 
correspond to 90\% upper limits. Filled symbols represent ellipticals ($T<-3$) and
open symbols represent lenticulars ($T>-3$).}
\end{inlinefigure}

\smallskip

\noindent
than cuspy galaxies. We searched the literature for HST derived
stellar surface brightness profiles of the galaxies in our sample, but only found 
results for 8 galaxies, so we cannot draw any general conclusions.  However,
the galaxies with the highest $L_{X,gas}$ (NGC4552) and lowest $L_{X,gas}$ (NGC3379)
in our sample both have flat cores (Lauer et al. 2005).  
There is a fairly strong correlation
between $L_{X,gas}$ and $kT$ among X-ray luminous early-type galaxies
(O'Sullivan, Ponman \& Collins 2003). Our sample of lower luminosity
galaxies, however, does not show such a correlation (see Fig. 6), suggesting that 
non-gravitational heating mechanisms have a more significant impact on the gas
properties of lower luminosity galaxies.

The gas density and mass can be estimated from the emission measure
of the best-fit MEKAL model to the diffuse emission along with an assumption 
about the spatial distribution of the gas.  The soft photon statistics in the Chandra 
data are insufficient to permit an empirical determination of the surface brightness 
profile of each galaxy.  Past studies have shown that the X-ray surface
brightness of luminous early-type galaxies is well fitted with
a $\beta$ model with $\beta \approx 0.5$ 
and a core radius between 1 and 3 kpc (Forman, Jones \& Tucker 1985).  
David, Forman \& Jones (1990,1991) generated a grid of elliptical galaxy models 
with different kinematic states of the hot gas (e.g., cooling flows, partial winds, 
total subsonic winds, and total transonic winds) and found that 
the surface brightness profile of these models could be characterized 
with $\beta$ between 0.3 and 0.5.  
Table 5 contains the derived central gas density and mass
in each galaxy assuming $\beta=0.5$ and a core radius
of 1~kpc.  This calculation also assumes that the diffuse emission 
arises from within a circular region centered on the galaxy with a radius 
of $R_m$ (given in Table 5) which has the same area as that in 
the region used to extract the X-ray spectrum of the diffuse emission.  
Varying $\beta$ between 0.3 and 0.67 or the core 
radius between 1 and 10~kpc only changes the estimated gas mass by 40\%.


All of the early-type galaxies in our sample are gas poor with 
an average 
$M_{gas} \approx 3 \times 10^7 M_{\odot}$ (see Table 5).  
Based on stellar 

\begin{inlinefigure}
\center{\includegraphics*[width=0.90\linewidth,bb=10 142 570 700,clip]{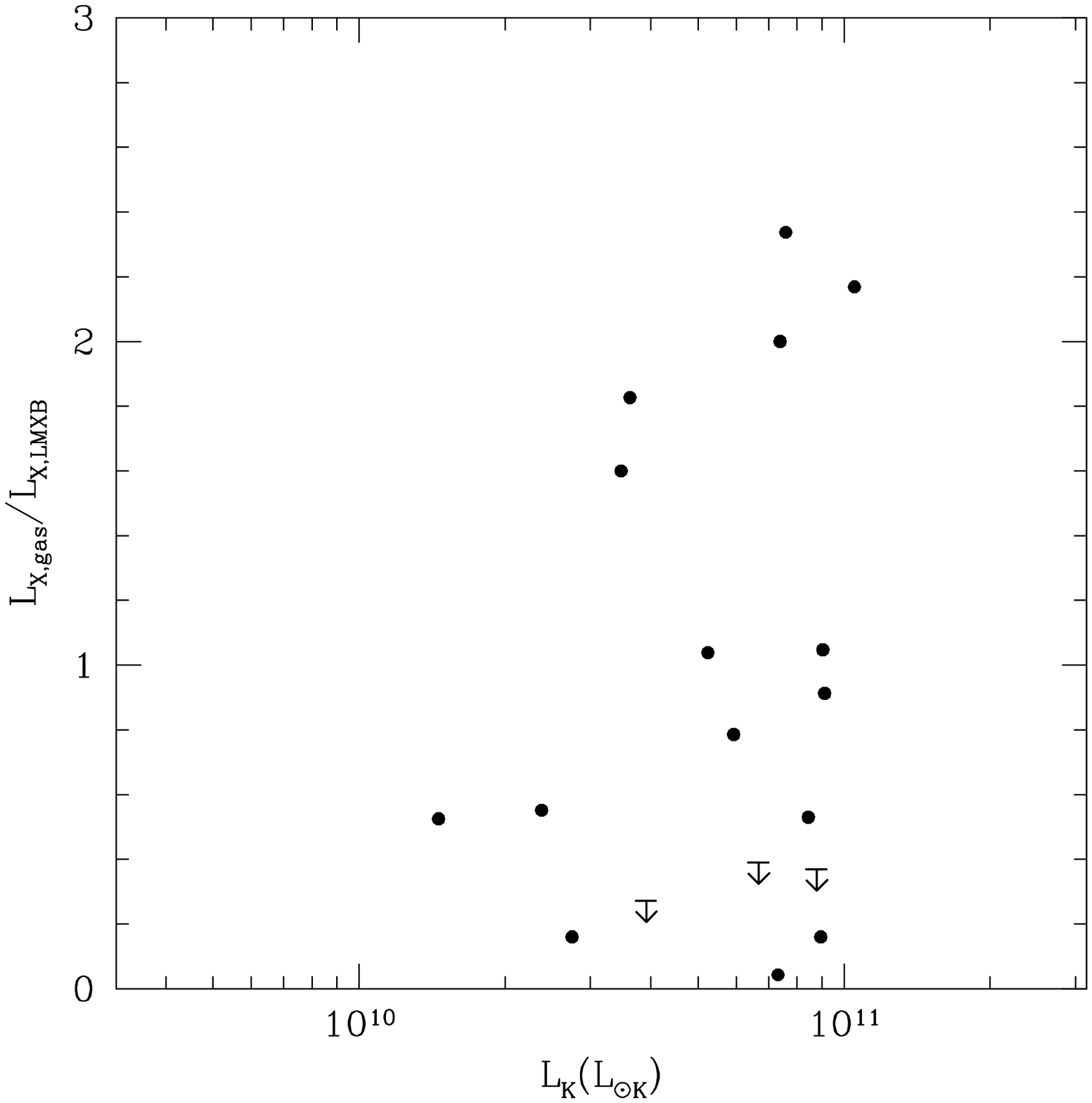}}  
\caption{Ratio of the gas luminosity to total LMXB luminosity vs. the K-band luminosity.}
\end{inlinefigure}

\smallskip

\noindent
evolution models of early-type galaxies, Bell \& de Jong (2001) 
derived an expression of $log(M_*/L_K) = -0.692+0.652(B-V)$,
where $M_*$ is the stellar mass.
The mean B-V in our sample is 0.9, which gives $M_*/L_K = 0.8 M_{\odot} / L_{\odot K}$ 
(consistent with dynamical measurements of a sample of early-type galaxies 
by Humphrey et al. 2006) and $M_{gas}/M_* = 8.0 \times 10^{-4}$.  A cumulative
histogram of $M_{gas}/M_*$ for our sample is shown in Fig. 7.
For comparison, Fukazawa et al. (2006) recently analyzed
Chandra observations of 53 X-ray luminous early-type galaxies and found
average values of $n_{e,0}= 0.1$~cm$^{-3}$, $M_{gas}= 3 \times 10^9 M_{\odot}$ 
and $M_{gas}/M_* = 0.01$.
Fig. 8 shows that there is significant scatter in $M_{gas}/M_*$ for a given
$L_K$ among the galaxies in our sample, but even the most gas rich system
in our sample (NGC4552) does not have as high a gas content as the mean value in the 
Fukazawa et al. sample.

For a Salpeter initial mass function, stars shed approximately 30\% 
of their initial mass over a Hubble time, which gives an 
average ejected mass of $2 \times 10^{10} \Mo$ per galaxy in our sample.
Very little of this gas can accrete into the central supermassive black
hole.  Based on the average stellar velocity dispersion
in our sample ($\sigma_* = 200$~km~s$^{-1}$) and the observed correlation between 
the central black hole mass and $\sigma_*$ (Gebhardt et al. 2000), the 
average $M_{bh}$ in our sample should be approximately $1.2 \times 10^{8} M_{\odot}$.  
To determine how long is takes to accumulate the observed gas mass, we use a
present day stellar mass loss rate from asymptotic giant branch
stars of $\dot M_* = 0.078 (L_B / 10^{10} L_{\odot~B}) M_{\odot}$~yr$^{-1}$, which
is consistent with the derived stellar mass loss rate from a sample of 9 ellipticals
observed by the {\it Infrared Space Observatory} (Athey et al. 2002).  
Converting this stellar mass loss rate to the K-band and using $L_K$ within 
the same aperture used to extract the diffuse spectrum (see Table 1), we find 
that the time to accumulate the derived gas mass in these galaxies is
$t_g \approx 1-3 \times 10^8$~yr (see Table 5).  
The typical cooling time of the gas 
($t_c = 5 k T M_{gas}/(2 \mu m_p L_{bol}))$ shown in Table 5 is 
$t_c \approx 10^9$~yr.  Since the cooling time is longer than 
the replenishment time, the gas cannot be condensing out of the hot phase and 
forming stars.  Thus, most of the gas shed by evolving stars must have been
expelled from these galaxies in a wind. 
The only exception to this is NGC4552 (M89), 

\begin{inlinefigure}
\center{\includegraphics*[width=0.90\linewidth,bb=10 142 570 700,clip]{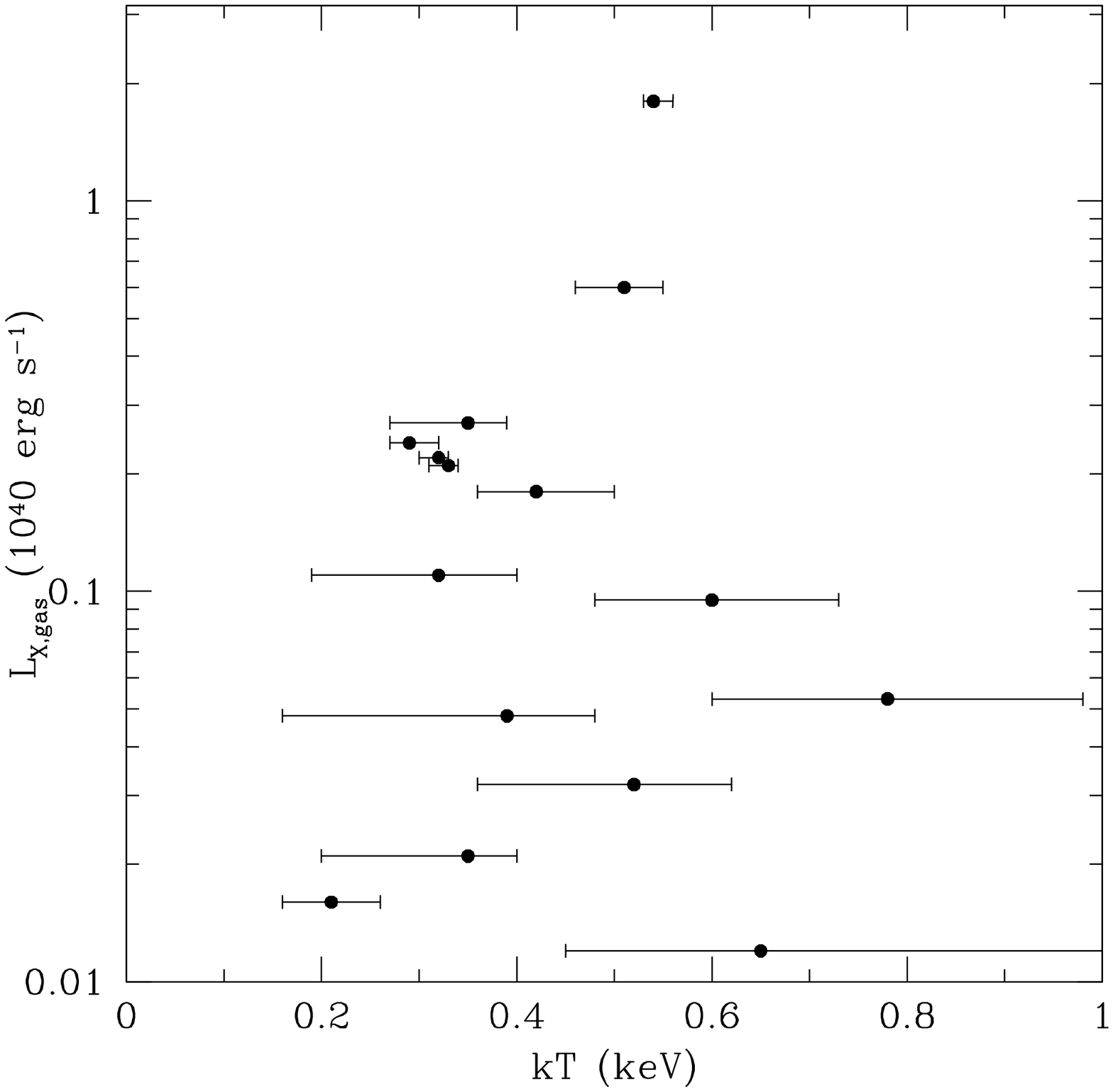}}  
\caption{Scatter plot of the 0.5-2.0~keV luminosity of the thermal component of the
diffuse emission ($L_{X,gas}$) against the best-fit temperature.  The errors
bars on the temperature are shown at the $1 \sigma$ confidence level.}
\end{inlinefigure}

\smallskip

\noindent
which is the most optically luminous 
galaxy in our sample, has the highest gas content, and is located 
in the densest environment. 
The Chandra image of NGC4552 shows that is has a significant gaseous
corona and that it is presently being 
ram pressure stripped of its gas as it falls into the Virgo cluster 
(Machacek et al. 2005).

There have been many theoretical studies regarding the dynamical evolution 
of the hot gas in early-type galaxies 
(e.g., Mathews \& Baker 1971; Loewenstein \& Mathews 1987; 
David, Forman \& Jones 1990,1991; Ciotti et al. 1991, 
Pellegrini \& Fabbiano 1994; Brighenti \& Mathews 1999).
The primary quantity that determines the present dynamic state of the gas 
is the ratio of the mass averaged temperature of the stellar ejecta
to the depth of the gravitational potential well of the galaxy.
The mass averaged temperature of stellar ejecta is given by
$T_* = \dot M_{SN Ia} T_{SN Ia}/ \dot M_*$, where 
$T_{SN Ia}=2 \mu m_p E_{SN Ia}/(3 k M_{ej})$ and $\dot M_{SN Ia}=M_{ej}\nu_{SN Ia}$.
Using $E_{SN Ia} = 10^{51}$~erg and $M_{ej}=1.4 M_{\odot}$, gives
$T_{SN Ia}=150$~keV per particle.  A large grid of models have been 
generated over the past 20 years covering a range of galaxy masses, SN Ia
rates, stellar initial mass functions and mass-to-light ratios.
Most models have used SN Ia rates between 1/4 and one times
Tammann's (1982) rate of $\nu_{SN Ia}=0.88 h^2$~SNU 
(where $h$ is the Hubble constant in units
of 100~km~s$^{-1}$~Mpc$^{-1}$ and a SNU = 1/100yr/$10^{10} L_{\odot B}$)
and a stellar mass loss rate of approximately 
$\dot M_* = 0.15 (L_B / 10^{10} L_{\odot~B}) M_{\odot}$~yr$^{-1}$
(Faber \& Gallagher 1976).
Converting to $H_0=70$~km~s$^{-1}$~Mpc$^{-1}$, these rates give
$T_* = 1.5-6.0$ keV per particle.
Using the more recent estimate of $\dot M_*$ from Athey et al. (2002)
and $\nu_{SN Ia}=0.16$~SNU (for $H_0=70$~km~s$^{-1}$~Mpc$^{-1}$) from 
Cappellaro et al. (1999), gives $T_* = 4.2$~keV per particle.
In the appendix, we show that the minimum energy required 
to remove the gas shed by stars in a wind is
equal to the mass averaged binding energy of the injected
gas (eq. 19).  We computed this minimum energy in detail
for NGC1023, which has a stellar velocity dispersion
close to the mean in our sample, assuming a Navarro, 
Frenk \& White (1996) distribution with a concentration 
parameter of 10 for the 

\begin{inlinefigure}
\center{\includegraphics*[width=0.90\linewidth,bb=10 142 570 700,clip]{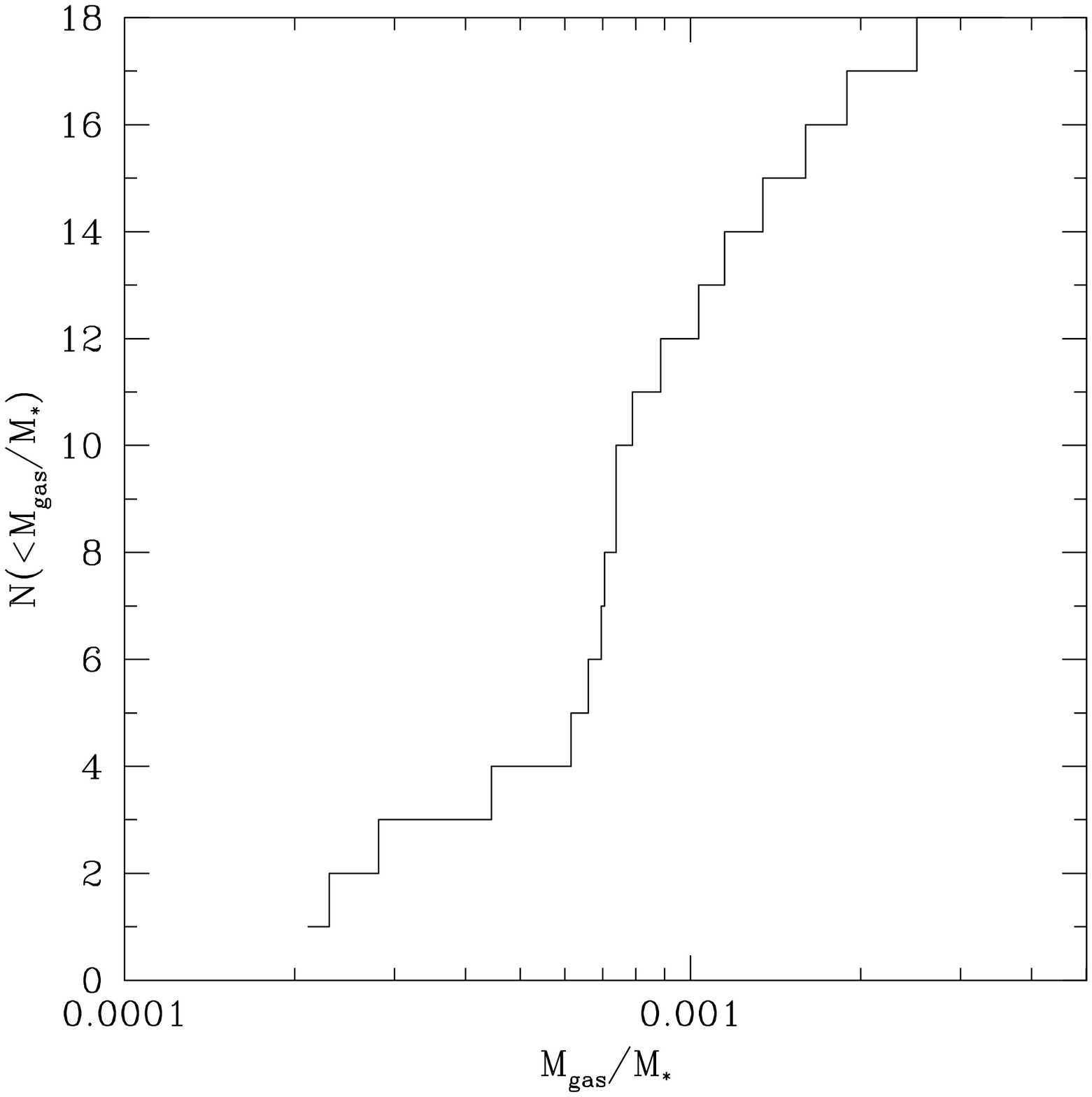}}  
\caption{Cumulative histogram of $M_{gas}/M_*$.}
\end{inlinefigure}

\smallskip

\noindent
dark matter and a King model for the 
stellar distribution with a core radius of $51.9^{\prime\prime}$
(Jarrett et al. 2003).  This calculation gives a minimum energy
of 2.01~keV per particle, which is equal to $7.2 \sigma_*^2$
in NGC1023.


If the gas is in a steady-state wind, 
then $\dot M_* = 4 \pi R_m^2 \rho_{gas}(R_m) u_w(R_m)$, 
where $\rho_{gas}(R_m)$ is the gas density at $R_m$, which can be estimated from the 
central gas density and the assumed $\beta$ model, and $u_w(R_m)$ is the
bulk gas velocity at $R_m$.  Using the stellar mass loss rates 
and central gas densities in Table 5 give $u_w \lax 35$~km~s$^{-1}$
(see Table 6).  The mechanical power of the winds 
($\dot E_w = \dot M_* u_w^2 / 2$) is a small fraction of the 
expected energy injection rate from SN Ia ($\dot E_{SNIa}$)
(see Table 6).  As shown in Fig.1, the gas in these galaxies
requires an additional heating of approximately 0.3~keV per particle above
purely gravitational heating, 
which is only 5\% of the expected SN Ia heating rate.
Supernova explosions in low-luminosity early-type galaxies should not experience 
significant radiative losses due to the lack of cold gas 
and the long cooling time of the hot gas.
Thus, most of the SN Ia supernova energy must be used in lifting
the gas out of the gravitational potential well of the galaxies.
Based on our detailed calculation for NGC1023 presented above,
the rate of increase in the total gravitational binding energy of the gas 
due to stellar mass loss can be estimated as 
$\dot W = 7.2 \dot M_* \sigma_*^2$. 
Table 6 shows that $\dot E_{SNIa}$ is 2-6 times $\dot W$ for the galaxies
in our sample based on the recent stellar mass loss rate in Athey et al. (2002)
and the SN Ia rate in Cappellaro et al. (1999), indicating that the energy released by SN Ia
is sufficient to drive galactic winds in these galaxies.
Using the Faber \& Jackson relation (1976), 
$\dot E_{SNIa}$/$\dot W \propto L_K^{-1/2}$, indicating
that heating by Type Ia supernovae is less efficient in driving the gas 
out of more luminous galaxies, which accounts for the higher gas content 
in these systems.

NGC 4552 is the only galaxy in our sample without a galactic wind.
NGC4552 and NGC3115 have the lowest ratios of $\dot E_{SNIa}$ to $\dot W$
in our sample with values of 2.0 and 1.9 (see Table 7).  Given the uncertainties
in $\dot E_{SNIa}$ and $\dot W$, it is possible that SN Ia are energetically
incapable of driving a wind in NGC4552.  While NGC4552 has the highest gas content
in our sample, it is still less than that found in more luminous early-type galaxies.

\begin{inlinefigure}
\center{\includegraphics*[width=0.90\linewidth,bb=10 142 570 700,clip]{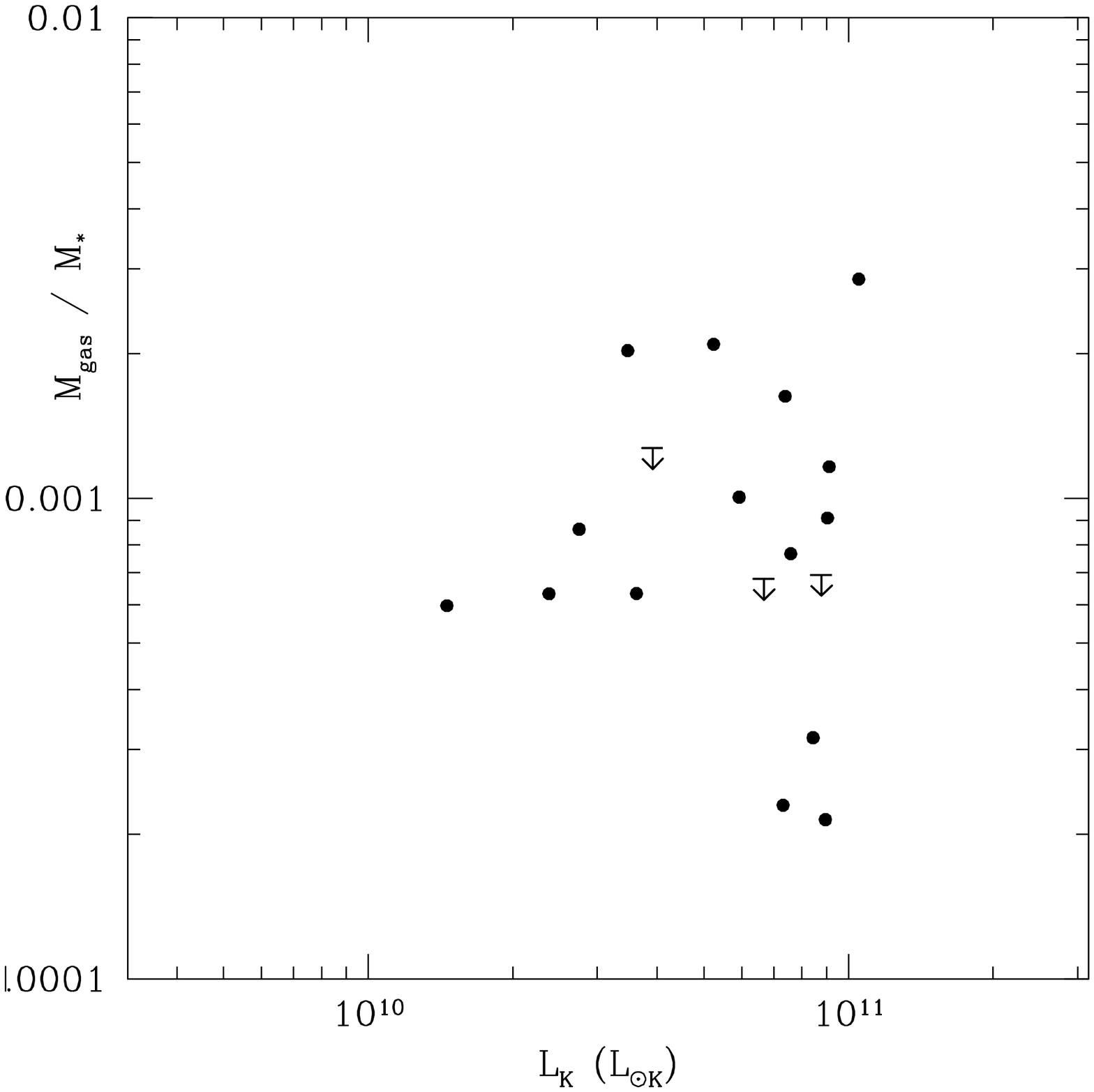}}  
\caption{Scatter plot of the ratio of gas mass to stellar mass ($M_{gas} / M_*$)
vs. the K-bank luminosity of the galaxies.}
\end{inlinefigure}

\smallskip

\noindent
The Chandra image shows that the low gas 
content in NGC4552 is due to ram pressure
stripping of its gas as it falls into the Virgo cluster.  Another
possible explanation for the lack of a recent wind and the build-up
of a gaseous corona in NGC4552 is that the ambient gas pressure in the Virgo 
cluster was sufficient to suppress the formation of a galactic wind before NGC4552 
experienced significant ram pressure stripping.  This would explain the low gas content 
of the more isolated galaxy NGC3115 (which was able to develop a galactic wind due to the
lack of a confining ambient medium) compared to NGC4552, even though they 
have very similar ratios of $\dot E_{SNIa}$ to $\dot W$.

\section{Implications for the Fe abundance of the Hot Gas}

The gas in these low luminosity galaxies should have an iron abundance
equal to the mass averaged abundance of the stellar and SN Ia
ejecta.  Based on our adopted rates given above, the gas should have an 
abundance 10 times the solar value in Grevesse \& Suval (1998) assuming 
that each SN Ia generates $0.7 \Mo$ of Fe.  
The primary difficulty in determining the gas density and abundance
(primarily Fe) of the hot gas in these galaxies is the difficulty in 
measuring the thermal continuum when there is residual emission from 
unresolved LMXBs in the diffuse emission.  Fig. 9 shows the
best-fit absorbed power-law plus MEKAL model (assuming solar
abundances) to the diffuse emission
in NGC1386, where 45\% of the total LMXBs emission is unresolved.
Notice that the power-law emission exceeds the thermal continuum
at all energies.  There is a strong degeneracy between the emission measure
and abundance in the fitting process.
For example, freezing the abundance at 10 times the solar value and refitting
the spectrum simply reduces the flux in the thermal continuum
by a factor of 10, while keeping the flux in the Fe-L lines
and the power-law component essentially the same.  Freezing the
normalization and index of the power-law model does not reduce
the degeneracy. The temperature determination is robust, since
it is derived from the centroid of the blended Fe-L lines.
Since the emission measure is proportional
to $n_e n_H$ and the flux in the Fe-L lines is proportional to
$n_e n_{Fe}$, it is difficult to accurately determine the
gas mass and Fe abundance in these galaxies due to 
the emission 

\begin{table*}[t]
\begin{center}
\caption{Hot Gas Energetics}
\begin{tabular}{lccccc}
\hline
Name & $u_w$ & $\dot E_w$ & $\dot E_{SN Ia}$ & $\dot W$ & $\dot E_{SN Ia} /\dot W$ \\
     & (km~s$^{-1}$) & (erg~s$^{-1}$) & (erg~s$^{-1}$) & (erg~s$^{-1}$) &  \\
     \hline\hline
ESO 0428-G014 & 8.9 & $1.6 \times 10^{36}$ & $4.3 \times 10^{40}$ & -  & - \\
NGC 821       & -   & -                    & $1.0 \times 10^{41}$ & $2.9 \times 10^{40}$ & 3.4 \\
NGC 1023      & 29  & $4.0 \times 10^{37}$ & $1.0 \times 10^{41}$ & $2.8 \times 10^{40}$ & 3.6 \\
NGC 1386      & 4.2 & $3.5 \times 10^{35}$ & $4.1 \times 10^{40}$ & $7.8 \times 10^{39}$ & 5.2 \\
NGC 1389      & -   & -                    & $4.6 \times 10^{40}$ & $5.6 \times 10^{39}$ & 8.2 \\
NGC 2434      & 6.4 & $1.8 \times 10^{36}$ & $8.7 \times 10^{40}$ & $2.4 \times 10^{40}$ & 3.6 \\
NGC 2787      & 6.8 & $3.9 \times 10^{35}$ & $1.7 \times 10^{40}$ & $4.5 \times 10^{39}$ & 3.8 \\
NGC 3115      & 35  & $6.2 \times 10^{37}$ & $1.0 \times 10^{41}$ & $5.2 \times 10^{40}$ & 1.9 \\
NGC 3245      & 10  & $3.5 \times 10^{36}$ & $7.0 \times 10^{40}$ & $2.4 \times 10^{40}$ & 2.9 \\
NGC 3377      & 11  & $2.0 \times 10^{36}$ & $3.2 \times 10^{40}$ & $3.8 \times 10^{39}$ & 8.4 \\
NGC 3379      & 31  & $2.8 \times 10^{37}$ & $8.6 \times 10^{40}$ & $2.4 \times 10^{40}$ & 3.6 \\
NGC 3608      & 6.0 & $1.1 \times 10^{36}$ & $6.2 \times 10^{40}$ & $1.7 \times 10^{40}$ & 3.6 \\
NGC 4251      & -   & -                    & $7.8 \times 10^{40}$ & $8.0 \times 10^{39}$ & 9.7 \\
NGC 4435      & 8.1 & $9.0 \times 10^{35}$ & $2.8 \times 10^{40}$ & $4.9 \times 10^{39}$ & 5.7 \\
NGC 4459      & 12  & $6.3 \times 10^{36}$ & $8.9 \times 10^{40}$ & $1.8 \times 10^{40}$ & 4.9 \\
NGC 4552      & 4.7 & $1.3 \times 10^{36}$ & $1.2 \times 10^{41}$ & $5.9 \times 10^{40}$ & 2.0\\
NGC 4697      & 11  & $6.6 \times 10^{36}$ & $1.1 \times 10^{41}$ & $2.0 \times 10^{40}$ & 5.5 \\
NGC 5866      & 11  & $6.4 \times 10^{36}$ & $1.1 \times 10^{41}$ & $1.9 \times 10^{40}$ & 5.8 \\
\hline
\end{tabular}
\end{center}

\noindent
Notes:  Galaxy name, required wind velocity ($u_w$) to expel the gas from 
the galaxy at the same rate as the present stellar mass loss rate, mechanical 
energy of the wind ($\dot E_w$), SN Ia heating rate ($\dot E_{SNIa}$), 
the required heating rate to remove the stellar mass loss from the gravitational
potential of the galaxy ($\dot W$) and the ratio $\dot E_w / \dot W$.
\end{table*}

\smallskip

\begin{inlinefigure}
\center{\includegraphics[angle=-90,width=3.5in,bb=50 50 700 800,clip]{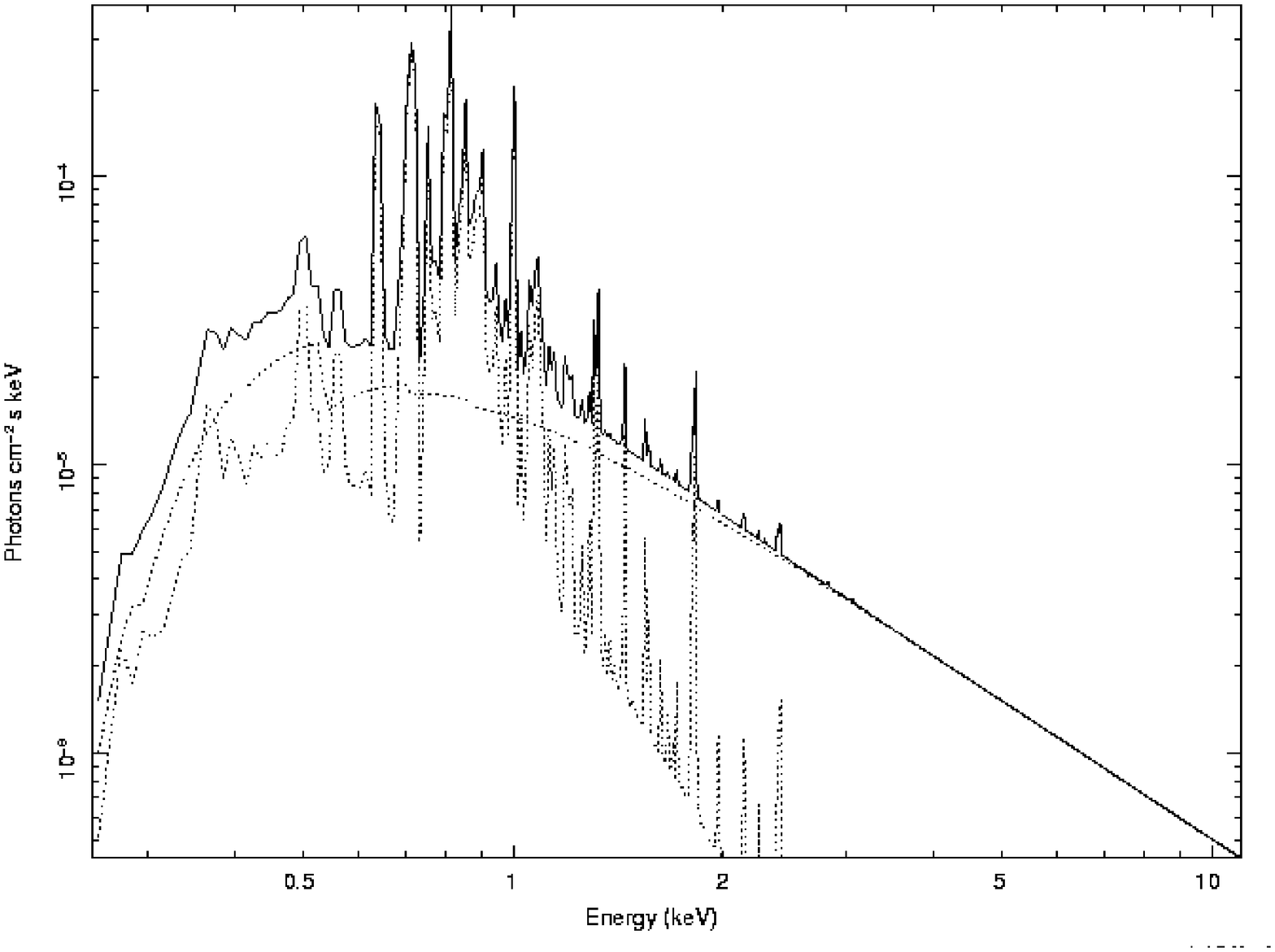}}  
\caption{Best-fit absorbed power-law plus MEKAL model to the
diffuse emission in N1386.  The lower curve shows the thermal emission,
the middle curve shows the power-law component, and the upper curve
is the sum of the thermal plus power-law models.}
\end{inlinefigure}

\smallskip

\noindent
from unresolved LMXBs.  If the gas in these galaxies has an 
Fe abundance that is 10 times the solar value, then the central gas densities, gas masses,
replenishment times and cooling times in Table 5 should be reduced by a factor of 3.  
In addition, the wind velocities in Table 6 should be increased by a factor of 3
and the mechanical power of the winds should by increased by an order of magnitude.
Hence, a higher Fe abundance actually strengthens the argument for galactic winds due 
to the decrease in gas mass and increase in wind velocity.
The conclusion that SN Ia are energetically capable of driving galactic
winds in these galaxies is not affected by the Fe abundance of the gas, since this 
argument mainly depends on the 
present SN Ia rate, stellar mass loss rate and the stellar velocity
dispersion in the galaxies.

\section{Is AGN Heating Important?}

Chandra has detected X-ray cavities 
and AGN driven shocks in several early-type galaxies (M84 - Finoguenov \& Jones 2002;
NGC4636 - Jones et al. 2002; Cen A - Kraft et al. 2003).  The
Chandra image of NGC4552 shows the presence of two spherical AGN driven shocks 
(Machacek et al. 2005).  None of the other galaxies in our sample
exhibit any substructure in the diffuse emission, but it is, of course,
much easier to detect cavities and shocks in galaxies with significant 
amounts of hot gas.
Seven of the galaxies in our sample were detected in the NVSS 
(Condon et al. 1998).  The 1.4 GHz luminosity ($\nu L_{\nu}$), or upper limit, 
for the galaxies is given in Table 7.  
Fig. 10 shows that there is no obvious trend between $M_{gas} / M_*$ 
and $\nu L_{\nu}$, however, there is a suggestion that 
the most radio luminous AGNs reside in galaxies with the highest gas content.

The radio luminosity of an AGN is a poor indicator of its mechanical
power. Using Chandra observations of a sample of galaxies and clusters
with X-ray cavities, Birzan et al. (2004) found that the ratio 
of mechanical to radio power varies from 10 in the most radio luminous 
AGNs, up to $10^4$ in systems with less radio luminous AGNs.  
Based on these results and a survey of radio-loud galaxies, Best et al. (2006) 
derived a time-averaged mechanical AGN heating rate of 
$1.6 \times 10^{41} (M_{bh}/ 10^8 M_{\odot})^{1.6}$~erg~s$^{-1}$.
Using the relation between black 

\begin{table*}[t]
\begin{center}
\caption{Radio Properties of Central AGN}
\begin{tabular}{lcc}
\hline
Name & $S_{\nu}$ & $\nu L_{\nu}$ \\
     & (mJy) & (erg~s$^{-1}$) \\
\hline\hline
ESO 0428-G014 & 81.0 & $8.0 \times 10^{37}$   \\
NGC 821       & $<2.3$ & $<2.2 \times 10^{36}$  \\
NGC 1023      & $<2.3$ & $<5.0 \times 10^{35}$  \\
NGC 1386      & 37.1 & $1.7 \times 10^{37}$   \\
NGC 1389      & $<2.3$ & $<1.8 \times 10^{36}$   \\
NGC 2434      & $<2.3$ & $<1.8 \times 10^{36}$  \\
NGC 2787      & 10.9 & $1.0 \times 10^{36}$  \\
NGC 3115      & $<2.3$ & $<3.6 \times 10^{35}$  \\
NGC 3245      &  6.7 & $4.9 \times 10^{36}$  \\
NGC 3377      & $<2.3$ & $<4.8 \times 10^{35}$  \\
NGC 3379      &  2.4 & $4.5 \times 10^{35}$  \\
NGC 3608      & $<2.3$ & $<2.0 \times 10^{36}$ \\
NGC 4251      & $<2.3$ & $<1.5 \times 10^{36}$ \\
NGC 4435      & $<2.3$ & $<5.0 \times 10^{35}$ \\
NGC 4459      & $<2.3$ & $<9.9 \times 10^{35}$  \\
NGC 4552      & 100.1 & $3.9 \times 10^{37}$ \\
NGC 4697      & $<2.3$ & $<5.2 \times 10^{35}$  \\
NGC 5866      &  21.8 & $8.5 \times 10^{36}$ \\

\hline
\end{tabular}
\end{center}

\noindent
Notes:  Galaxy name, 1.4~GHz flux ($S_{mJy}$) and luminosity ($\nu L_{\nu}$) of
the central AGN.
\end{table*}

\smallskip

\begin{inlinefigure}
\center{\includegraphics*[width=0.90\linewidth,bb=10 142 570 700,clip]{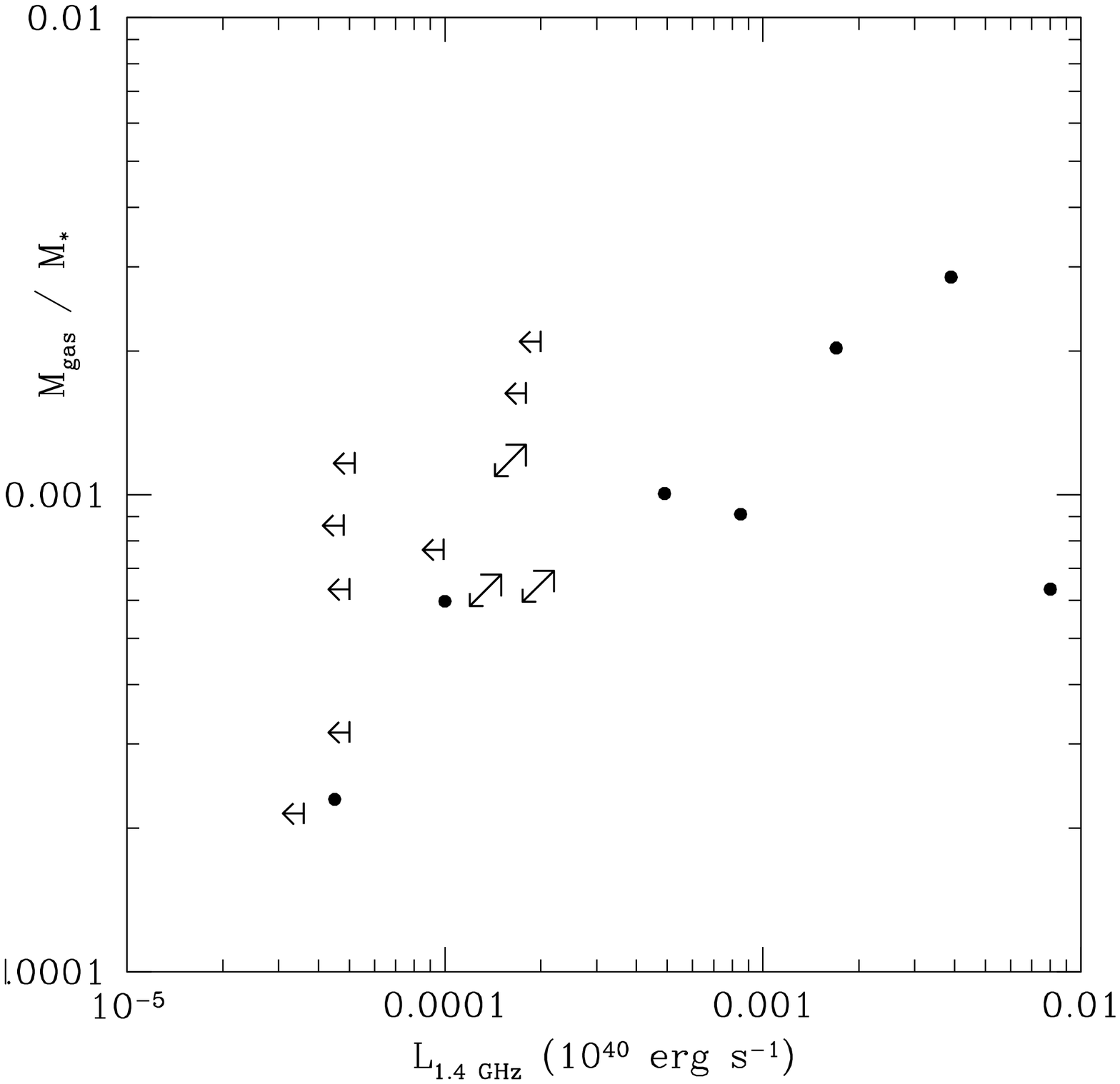}}  
\caption{Scatter plot of the gas mass to stellar mass ratio ($M_{gas} / M_*$)
vs. the 1.4~Ghz radio luminosity of the central AGN. The leftward pointing arrows
represent upper limits on the 1.4~GHz luminosity and the arrows pointing to the 
lower left represent upper limits on both the 1.4~GHz luminosity and
gas mass.}
\end{inlinefigure}

\smallskip
\noindent
hole mass and the absolute red magnitude,
$M_R$, found by
McLure \& Dunlop (2002) of $log(M_{bh}/M_{\odot}) = -0.50M_R-2.96$,
and an average $R-K$ of 2.5 for the galaxies in our sample, we plot the expected 
AGN and SN Ia heating rates in Fig. 11 against the bolometric luminosity
of the hot gas.
For comparison, the mechanical power required to generate the observed
shocks in NGC4552 (the upper right hand point in Fig. 11) 
is $3 \times 10^{41}$~erg~s$^{-1}$ (Machacek et al. 2005), which is
a factor of 2 greater than the rate predicted by Best et al. (2006),
but certainly within the uncertainties. Fig. 11 indicates that 
heating by SN Ia should dominate the energetics of
the gas in early-type galaxies with $L_K \lax 10^{11} L_{\odot K}$ and that 
AGN heating should dominate in more luminous galaxies.

\noindent

\noindent

\section{Discussion}

If heating by SN Ia is the dominant mechanism for 
expelling the gas shed by evolving stars from low-luminosity 
early-type galaxies, it is a puzzle as to why there is so much 
scatter in $M_{gas} / M_*$ and $L_{X,gas}$ (see Figs. 7 and 11) given 
that supernova heating is essentially a continuous process.  
Since the gas masses are computed in different apertures (see Table 5),
we checked to see if there is a correlation between $M_{gas} / M_*$
and aperture size and found none.  While the time-averaged 
SN Ia heating rate shown in Fig. 11 exceeds the time-averaged AGN heating rate,
except for possibly the most luminous galaxies in our sample, the 
episodic nature of AGN outbursts could produce times 
during which mechanical AGN heating exceeds SN Ia heating.
Thus, the large observed scatter could be a reflection of 
recent AGN activity.  The sound crossing time within the central
10~kpc in these galaxies is approximately $3 \times 10^7$~yr.  
If AGN outbursts repeat on a shorter time scale, then the gas 
will not be able to establish a steady-state wind. 
In the very center of the 
galaxies, heating by SN Ia
cannot be thought of as a continuous process.  For example, within
an enclosed stellar mass of $10^6 M_{\odot}$, there will only be
one SN Ia every $3.0 \times 10^7$~yr.  During this time, the stars within 
this region will shed a total of 50~$M_{\odot}$ of gas.  If this gas is converted 
into mechanical energy by the central black hole with an efficiency of 10\%, this 
will produce $1.0 \times 10^{55}$~erg of energy at a rate of 
$1.0 \times 10^{40}$~erg~s$^{-1}$, which 
is comparable to the SN Ia heating rate of an entire galaxy (see Fig. 10). 
Thus, it could be that the gas in the very center of the 

\begin{inlinefigure}
\center{\includegraphics*[width=0.90\linewidth,bb=10 142 570 700,clip]{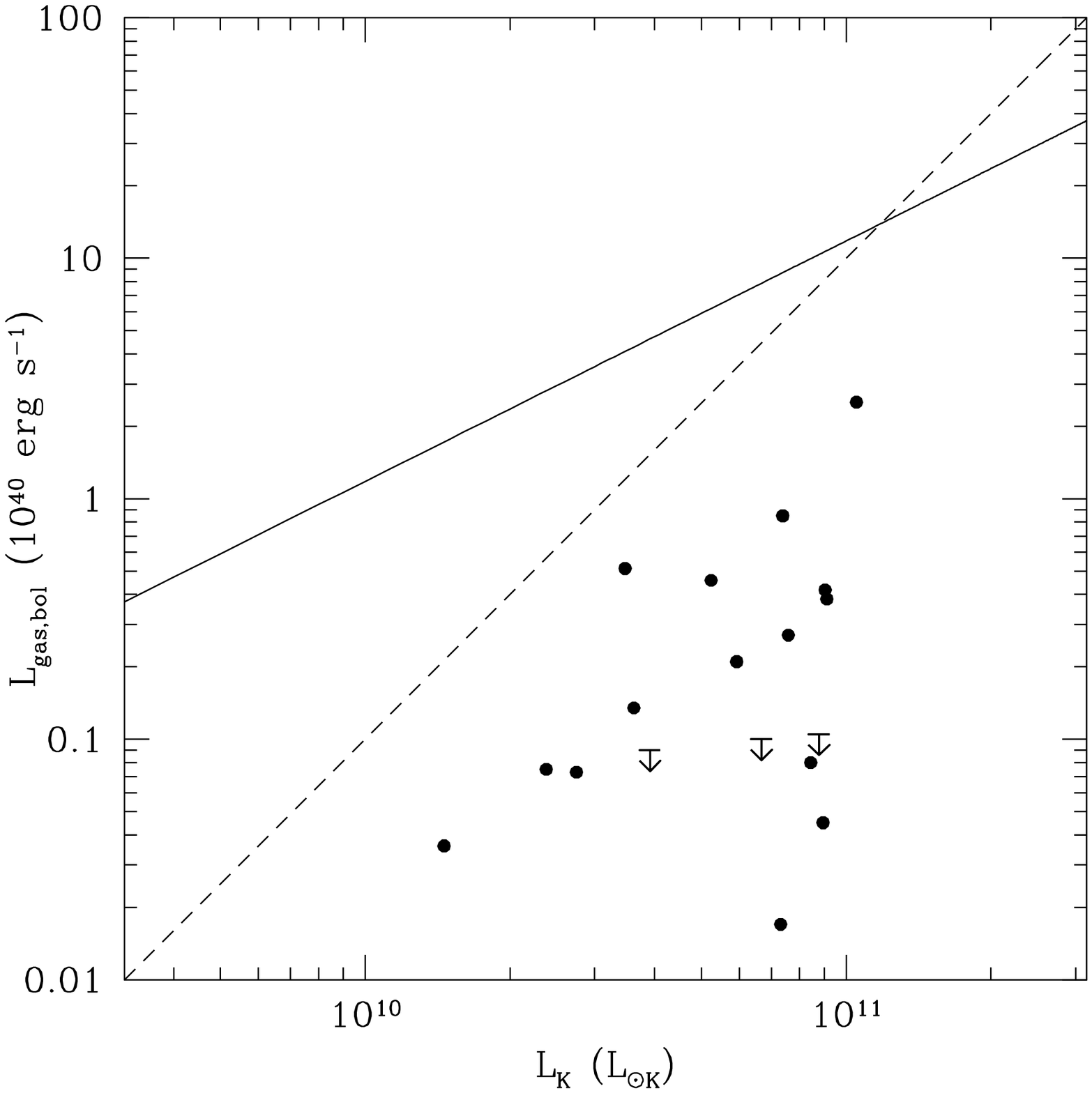}}  
\caption{Scatter plot of the bolometric X-ray luminosity of the gas vs. the K-band luminosity of 
the galaxies. The inverted arrows are 90\% upper limits. The solid line is the estimated 
SN Ia heating rate and the 
dashed line is the estimated mechanical AGN heating rate.}
\end{inlinefigure}

\smallskip

\noindent
galaxies does not partake 
in the global SN Ia wind, but periodically accretes into the central black hole.


Environment can also play a significant role in the gas content
of early-type galaxies. XMM-Newton and Chandra observations of 
early-type galaxies in the Coma cluster have shown that these galaxies 
can be either under luminous or over luminous in X-rays relative to 
field galaxies with comparable optical luminosities due to the
competing effects of adiabatic compression, ram pressure stripping
and thermal evaporation
(Vikhlinin et al. 2001; Finoguenov \& Miniati 2004; 
Hornschemeier et al. 2005).  Due to the large distance of the Coma cluster,
these observations were only sensitive to the X-ray properties of luminous
early-type galaxies which have massive hydrostatic coronae in regions 
outside of dense groups or clusters.
Low luminosity field galaxies, however, should have strong winds due to 
the lack of a confining ambient medium. 


NGC4552 is located in the densest environment of any of the galaxies in 
our sample and clearly shows the effects of its environment. 
It is located $72^{\prime}$ (a projected distance of 360~kpc) from the 
center of the Virgo cluster and has the highest gas content and hosts the 
most radio luminous AGN in our sample.  
The Chandra image shows that NGC 4552 has a leading
cold front and a trailing wake of ram pressure stripped gas 
(Machacek et al. 2005).  Based on the pressure jump across the cold front, 
Machacek et al. estimated that NGC4552 is traveling at 1600~km~s$^{-1}$.  
The average gas density and pressure in NGC4552 derived from the values in 
Table 5 are $<n_e> = 1.7 \times 10^{-3}$~cm$^{-3}$ and 
$<P>=1.9 \times 10^{-3}$~keV~cm$^{-3}$.  Using the results in
Machacek et al., the ram pressure of the Virgo gas at the leading cold
front in NGC4552 is $P_{ram} = 2.2 \times 10^{-3}$~keV~cm$^{-3}$, which 
is greater than the average thermal pressure of the gas, but less than 
the central pressure of $P_0 = 3.5 \times 10^{-2}$~keV~cm$^{-3}$.  These calculations
are in good agreement with the fact that some of the outer gas in NGC452 is 
being stripped as it falls into the Virgo cluster, but not the central gas.
Based on the inferred velocity of NGC4552, it should  
have traveled approximately 2~Mpc during the time it has taken to 
accumulate its present gas mass.  The ROSAT all-sky survey data
show that the gas in Virgo follows a $\beta=0.47$ profile (Schindler et al 1999).  
Assuming the Virgo gas is isothermal and a radial trajectory for NGC4552,
the thermal pressure of the Virgo gas surrounding NGC4552 would have increased 
by a factor of 20 during this time and the ram pressure by an even
greater factor due to the increasing infall velocity.  
Thus, the ambient pressure
during most of this time was sufficient to suppress the formation of 
a galactic wind, but insufficient to strip the gas from the galaxy. This 
would have lead to a substantial increase in gas mass which could have 
triggered the AGN outburst.

\section{Summary}

We have presented a systematic analysis of Chandra observations 
of 18 low-luminosity early-type galaxies.  Emission from LMXBs is very 
significant in these galaxies
and comprises between 30 to 95\% of the total 0.5-2.0~keV emission. 
We find that the combined X-ray luminosity of LMXBs (resolved plus 
the power-law component of the unresolved emission) scales approximately 
linearly with $L_K$ among the galaxies in our sample with an average 
$L_{X,LMXB} / L_K$ ratio comparable to that found in more luminous 
early-type galaxies.  The gas temperature in these galaxies varies 
from 0.2 to 0.8~keV and roughly follows the observed trend between 
gas temperature and stellar velocity dispersion 
among more luminous early-type galaxies, albeit with significant scatter.
The ratio of energy per unit mass in the galaxies to that in the gas, 
$\beta_{spec}$, varies from 0.3 to 1.0, indicating the presence of significant
non-gravitational heating in most of the galaxies.
We do not find any correlation between gas temperature and $L_{X,gas}$, unlike more 
luminous galaxies.

All of the galaxies in our sample are gas poor with 
$M_{gas} / M_* = 0.2-3.0 \times 10^{-3}$, compared to 
$M_{gas} / M_* = 0.01$ in more luminous early-type galaxies.  
The observed gas mass is much less than that expected from the accumulation
of stellar mass loss over the lifetime of the galaxies.  Based on recent 
estimates of the stellar mass loss rate in early-type galaxies, the time 
required to accumulate the observed gas mass is $10^8 - 10^9$~yr.
Due to the low gas densities, the cooling time of the gas is significantly longer 
that the replenishment time.  Thus, most of the gas shed by evolving stars 
in these galaxies could not have condensed out of the hot phase and 
formed stars, and must have been expelled from the galaxies in a wind.
The only exception to this is the most optically luminous galaxy in our
sample, NGC4552.

Based on recent estimates of the SN Ia rate and mechanical 
AGN heating rate in early-type galaxies, we find that, on average, heating by 
type Ia supernova should dominate over AGN heating in galaxies with
$L_K \lax 10^{11} L_{\odot K}$.  Using the stellar mass loss rate in 
Athey et al. (2002) and the SN Ia rate in Cappellaro et al. 
(1999), we find that SN Ia are energetically capable of driving
galactic winds in low-luminosity early-type galaxies even if the present 
SN Ia rate has been overestimated by a factor of two or the 
stellar mass loss rate has been underestimated by a factor of two.
We also find that most of the supernova energy must be consumed 
in lifting the gas out of the gravitational potential well of
the galaxies, with little energy converted into thermal
or bulk kinetic energy of the gas.

If heating by SN Ia were the dominant heating mechanism at
all times, it would be difficult to account for the large scatter
in gas properties among the galaxies in our sample.
Even though our calculations show that the time-averaged AGN heating 
rate is less than the supernova heating rate, the episodic nature
of AGN outbursts could produce periods when AGN heating dominates.
One possible explanation for the large scatter in gas properties
is that galaxies with very little gas at the present time
recently experienced a significant AGN outburst that produced 
a rapid expulsion of the gas.  Six of the galaxies in our sample 
were detected in the NVSS (Condon et al. 1998), however, we do 
not find any correlations between $M_{gas} / M_*$ and radio power.  
The lack of a correlation may simply result from the time delay
between when the AGN reaches its peak radio power and when the gas is
expelled from the galaxy.  In fact, the most radio luminous 
AGN in our sample has the largest gas mass.
Another possible source for the observed scatter in gas properties
among early-type galaxies is variations in the ambient gas pressure surrounding 
the galaxies which would impact their ability to develop galactic winds. 

\medskip

We would like to thank the referee, Mike Loewenstein, for his many useful
comments. We also had many useful discussions of our work with Ewan O'Sullivan
and Jan Vrtilek. This work was supported by NASA grant AR4-5016X.



\newpage

\appendix
\section*{Wind Sonic Point}

\newcommand{\mdot}{\dot M}
\newcommand{\half}{{1\over 2}}
\newcommand{\alphabar}{\overline{\alpha}}
\newcommand{\vk}{v_{\rm K}}
\newcommand{\rs}{r_{\rm s}}
\newcommand{\phibar}{\overline{\phi}}
\newcommand{\emin}{\epsilon_{\rm min}}
\newcommand{\rcut}{r_{\rm c}}

Consider a steady, spherical wind driven by energy injected in
association with stellar mass loss.  Following Mathews \& Baker (1971), 
we assume
that stellar mass loss is distributed smoothly and mixes well with the
hot ISM, so that we may treat the wind as a single, smooth fluid, with
distributed mass injection.  The equation of mass conservation is then
\begin{equation}
{1\over r^2} {d \over dr} \rho v  r^2 = \alpha,
\end{equation}
where $r$ is the radius, $\rho$ is the gas density, $v$ is its radial
velocity and $\alpha(r)$ is the mass injection rate per unit volume
due to stellar mass loss.  Assuming that the wind extends right to the
center of the galaxy, this equation can be integrated to give the mass
flow rate through a sphere of radius $r$,
\begin{equation}
\mdot(r) = 4\pi \rho v r^2 = \int_0^r \alpha(r') 4 \pi r'^2 \,
dr' = {4\pi\over 3} \alphabar r^3, \label{eqn:mass}
\end{equation}
where we have also defined $\alphabar(r)$, the mean stellar mass loss
rate per unit volume within $r$.  The momentum equation for the steady
wind is
\begin{equation}
\rho v {dv\over dr} = - {dp\over dr} - \rho g - \alpha v,
\label{eqn:mom} 
\end{equation}
where $p$ is the gas pressure and $g(r)$ is the acceleration due to
gravity (positive inward).  The energy equation may be written
\begin{equation}
{1\over r^2} {d \over dr} \rho v r^2 (H + \half v^2 + \phi)
= \alpha (\epsilon + \phi),
\end{equation}
where $H$ is the specific enthalpy, 
\begin{equation}
H = \gamma p / [(\gamma - 1) \rho],  \label{eqn:enthalpy}
\end{equation}
$\gamma$ is the ratio of specific heats, $\phi$ is the gravitational
potential and $\epsilon$ is the total energy per unit mass that is
deposited in the flow along with the stellar mass loss.  This is due
principally to supernova heating, but includes kinetic energy due to
random stellar motions, energy injected by planetary nebulae, etc.
Following the usual practice, we assume that $\epsilon$ does not vary
throughout the galaxy.  Integrating the energy equation gives
\begin{equation}
\mdot (H + \half v^2 + \phi) = \int_0^r \alpha(r') [\epsilon +
  \phi(r')] 4 \pi r'^2 \, dr' = \mdot w, \label{eqn:wdef}
\end{equation}
which also defines $w(r)$, a known function of the radius, and we have
\begin{equation}
w(r) = H + \half v^2 + \phi. \label{eqn:bernoulli}
\end{equation}
We ignore the effects of radiative energy loss, which would clearly
increase the energy required to drive the wind.  These are generally
small for the systems considered here.

From equation (\ref{eqn:mass}), the density can be given in terms of
the velocity as
\begin{equation}
\rho = {\alphabar r \over 3 v},
\end{equation}
and from equations (\ref{eqn:bernoulli}) and (\ref{eqn:enthalpy}) we
get 
\begin{equation}
{p\over\rho} = {\gamma - 1\over \gamma} (w - \half v^2 -
\phi). \label{eqn:temp} 
\end{equation}
These can be used to eliminate the pressure derivative from the
momentum equation (\ref{eqn:mom}), giving, after some algebra,
\begin{equation}
\rho (v^2 - s^2) {dv\over dr} = {\alphabar \over3} \left[ 2 (\gamma -
  1) (w - \half v^2 - \phi) - \vk^2 \right] - \alpha \left[ {\gamma +
    1\over 2} v^2 + (\gamma - 1) \epsilon \right], \label{eqn:spmom}
\end{equation}
where the sound speed, $s$, is given by 
\begin{equation}
s^2 = \gamma p /\rho \label{eqn:sound}
\end{equation} 
and the Kepler speed, $\vk$, is defined by $\vk^2(r) = g r$.  Note
that in order to obtain this result, we have used $g = d\phi/dr$ and
differentiated the definitions of $\alphabar$ (equation
\ref{eqn:mass}) and $w$ (equation \ref{eqn:wdef}). 

The wind solutions starts with $v=0$ at $r=0$, passes through a sonic
point, where $v = s$, and ends up as a freely expanding, supersonic
flow for $r\to \infty$.  At the sonic point $v^2 = s^2 = (\gamma - 1)
(w - v^2/2 - \phi)$ (equations \ref{eqn:temp} and \ref{eqn:sound}), so
that
\begin{equation}
v^2(\rs) = {2(\gamma - 1) \over \gamma + 1} [w(\rs) - \phi(\rs)],
\label{eqn:atsp}
\end{equation}
where $\rs$ is the radius of the sonic point.  In order for the
solution to pass through the sonic point, the right hand side of
equation (\ref{eqn:spmom}) must vanish there.  After using equation
(\ref{eqn:atsp}) to eliminate the velocity, the resulting condition
may be written
\begin{equation}
\left[ {4 \alphabar(\rs) \over 3 (\gamma + 1)} - \alpha(\rs) \right]
     [w(\rs) - \phi(\rs)] = \alpha(\rs) \epsilon + {\alphabar
     \vk^2 \over 3 (\gamma - 1)}.  \label{eqn:sola}
\end{equation}
The terms on the right of this equation are both positive and, since
$H + v^2/2 > 0$, equation (\ref{eqn:bernoulli}) requires $w - \phi>0$
too.  Thus
\begin{equation}
{4 \alphabar(\rs) \over 3 (\gamma + 1)} > \alpha(\rs). \label{eqn:ineq}
\end{equation}
Defining
\begin{equation}
\phibar(r) = {\int_0^r \alpha(r') \phi(r') r'^2 \, dr' \over \int_0^r
  \alpha(r') r'^2 \, dr'}, \label{eqn:pbdef}
\end{equation}
from the definition of $w$ (equation \ref{eqn:wdef}), we have $w(r) =
\epsilon + \phibar(r)$.  Using this in equation (\ref{eqn:sola}) and
solving for $\epsilon$ gives 
\begin{equation}
\epsilon \left[ {4\alphabar(\rs) \over 3 (\gamma + 1)} - 2 \alpha(\rs)
    \right] = \left[ {4 \alphabar(\rs) \over 3 (\gamma + 1)} -
    \alpha(\rs) \right] [\phi(\rs) - \phibar(\rs)] + {\alphabar(\rs)
    \over 3 (\gamma - 1)} \vk^2(\rs).
\end{equation}
Since $\phi$ is an increasing function of the radius, $\phi(\rs) -
\phibar(\rs) > 0$, and we showed that the factor multiplying it is
positive above (\ref{eqn:ineq}).  Thus, we must also have
\begin{equation}
{2 \alphabar(\rs) \over 3 (\gamma + 1)} > \alpha(\rs).
\label{eqn:large}
\end{equation}
Under the usual assumption that $\alpha$ is proportional to the
stellar density, this condition requires the mean density to be
substantially greater than the local density at the sonic point.
Thus, the sonic point must occur well outside the dense stellar core
of a typical galaxy.  We can now rewrite the result for $\epsilon$ as 
\begin{equation}
\epsilon = \left[ \phi - \phibar + {(\gamma + 1) \vk^2 \over 4
  (\gamma - 1)} + \alpha \left\{ \phi - \phibar + {(\gamma + 1)
  \vk^2 \over 2 (\gamma - 1)} \right\} \Bigg / \left\{ {4
  \alphabar \over 3 (\gamma + 1)} - 2 \alpha \right\} \right]_{r =
  \rs}, \label{eqn:result}
\end{equation}
knowing that all of the factors in the term multiplied by $\alpha$ are
positive.  For a galaxy model with specified potential and stellar
mass loss rate, $\alpha$, this expression enables us to determine the
specific energy injection rate, $\epsilon$, from the location of the
sonic point.  Alternatively, if the energy injection rate is known,
this equation may be solved to find the location of the sonic point.
In particular, in the limit of large energy input, the sonic point
approaches the radius that makes the denominator of the second term
vanish, i.e.,  that makes the inequality (\ref{eqn:large}) an indentity.

For the remainder of this discussion, we assume that the stellar mass
loss rate is proportional to the stellar density.  Inside a galaxy,
the right hand side of the expression for $\epsilon$ is generally a
decreasing function of the sonic radius, $\rs$, so that the energy
input required to drive a wind is minimized when the sonic point
occurs at large $r$.  If the stellar density falls rapidly to zero at
the edge of the galaxy, $\alpha$ falls to zero quickly, while the
other terms in this expression are continuous.  Thus there is a rapid
reduction in the energy required to drive the wind as the sonic point
moves across the outer edge of the galaxy.  As a result, the sonic
point occurs close to the edge of the stellar distribution for a range
of values of $\epsilon$.

Beyond the edge of the stars $\alpha = 0$ so that $\phibar(r)$ is
constant (from the definition \ref{eqn:pbdef}).  The remaining
variable terms in (\ref{eqn:result}), $\phi(r) + (\gamma + 1) \vk^2(r)
/ [4 (\gamma - 1)]$, may increase or decrease with radius, depending
on the value of $\gamma$ and the distribution of the gravitating
matter (which may extend beyond the stars).  For $\gamma=5/3$, these
terms reduce to $\phi(r) + \vk^2(r)$, which is a non-decreasing
function of $r$ (since $\vk^2(r) = GM(r)/r$, where $M(r)$ is the
gravitating mass within $r$), so that the global minimum of the
expression (\ref{eqn:result}) occurs at a value of $\rs$ close to, or
immediately outside the edge of the stellar distribution.  However,
the local condition for the sonic point (\ref{eqn:result}) does not
guarantee that the wind solution can be extended to infinity.  Beyond
the edge of the stars, $w(r)$ is also constant, and it is evident from
equation (\ref{eqn:bernoulli}) that it necessary that $w = \epsilon +
\phibar \ge0$ for the wind solution to extend to $r=\infty$, an
essential requirement for a steady wind.  Thus the global minimum
value of $\epsilon$ required for a wind when $\gamma = 5/3$ is
\begin{equation}
\emin = -\phibar(\infty).
\end{equation}
Similar considerations apply for $\gamma < 5/3$, except that the
energy requirement may now be minimized for $\rs \to \infty$ in
equation (\ref{eqn:result}).  Thus we arrive at the same global value
for $\emin$.  Note that, for $\gamma = 5/3$, the wind solution with $w
= 0$ attains a constant Mach number beyond the edge of the gravitating
matter, rather than having the Mach number approach infinity.  For
$\gamma < 5/3$, when the energy requirement is minimized for $\rs\to
\infty$, the flow would be subsonic everywhere.  In that case, the
effect of a disturbance at finite $r$ is felt throughout the wind
flow.  In either case, the wind solution is fragile.  This is not
surprising for a wind driven with the absolute minimum energy.  It
emphasizes that the minimum energy requirement is just that.  In
reality, greater energy is required to sustain a wind.

To evaluate $\phibar$, we assume that the gravitational potential has
the NFW form,
\begin{equation}
\phi(r) = - 4 \pi G \rho_0 a^2 \left[ {\ln (1 + r/a) \over r/a } -
  {1 \over 1 + c} \right], \qquad\qquad {\rm for\ } r < ac,
\end{equation}
where $a$ is the scale length, $\rho_0$ is the scale density and $c$
is the concentration parameter.  The stars, hence the mass injection
rate, are assumed to have the King model distribution
\begin{equation}
\alpha(r) = {\alpha_0 \over (1 + r^2/b^2)^{3/2}},
\end{equation}
where $b$ is the core radius.  Using the definition (\ref{eqn:pbdef}),
we then get 
\begin{eqnarray}
\phibar(r) = - 4 \pi G \rho_0 a^2 \left[-{1\over 1 + c} + {1
    \over \ln (y + \sqrt{1 + y^2}) - y / \sqrt{1 + y^2}} \right.
\nonumber \\
\left. \times \left\{ - {\ln (1 +
    \beta y) \over \beta \sqrt{1 + y^2}} + {1\over \sqrt{1 + \beta^2}}
    \ln {(1 + \beta y + \beta \sqrt{1 + y^2} - \sqrt{1 + \beta^2}) (1
    + \beta + \sqrt{1 + \beta^2}) \over (1 + \beta y + \beta \sqrt{1 +
    y^2} + \sqrt{1 + \beta^2}) (1 + \beta - \sqrt{1 + \beta^2})}
    \right\} \right], \\
\end{eqnarray}
where $\beta = b/a$, $y = \rcut/b$ and the stellar distribution is cut
off at $\rcut$.  The normalizing factor for the NFW potential, $4\pi G
\rho_0 a^2$, was determined from the assumption that the maximum
rotation speed is related to the line-of-sight velocity dispersion by
$v_{\rm K,max}^2 = 2 \sigma^2$, giving
\begin{equation}
4 \pi G \rho_0 a^2 = \lambda \sigma^2,
\end{equation}
with $\lambda \simeq 9.25$.  Equating the mean density of the halo to
$\chi$ times the critical density then gives the scale length,
\begin{equation}
a = {\sigma \over H_0} \sqrt{2 \lambda [\ln (1 + c) - c/(1 + c)] \over
  \chi c^3},
\end{equation}
where $H_0$ is the Hubble constant.  Parameters for the King model
are set to match photometric properties of the galaxies.

\newpage

\end{document}